\documentstyle[amstex,amssymb,preprint,aps]{revtex}

\begin{document}
\title{Collective Modes of Quantum Hall Stripes}
\author{R. C\^{o}t\'{e}$^{1}$ and H.A. Fertig$^{2}$}
\address{$^{1}$ D\'{e}partement de Physique, Universit\'{e} de Sherbrooke, \\
Sherbrooke, Qu\'{e}bec, Canada J1K-2R1}
\address{$^{2}$ Department of Physics and Astronomy, University of Kentucky,\\
Lexington KY 40506-0055}
\date{\today}
\maketitle

\begin{abstract}
The collective modes of striped phases in a quantum Hall system are computed
using the time-dependent Hartree-Fock approximation. Uniform stripe phases
are shown to be unstable to the formation of modulations along the stripes,
so that within the Hartree-Fock approximation the groundstate is a stripe 
{\it crystal}. Such crystalline states are generically gapped at any finite
wavevector; however, in the quantum Hall system the interactions of
modulations among different stripes is found to be remarkably weak, leading
to an infinite collection of collective modes with immeasurably small gaps.
The resulting long wavelength behavior is derivable from an elastic theory
for smectic liquid crystals. Collective modes for the phonon branch are
computed throughout the Brillouin zone, as are spin wave and magnetoplasmon
modes. A soft mode in the phonon spectrum is identified for partial filling
factors sufficiently far from $1/2$, indicating a second order phase
transition. The modes contain several other signatures that should be
experimentally observable.

\end{abstract}

\section{Introduction and Summary of Results}

Recently, it has been discovered\cite{lilly,du} that high quality
two-dimensional electron systems in the quantum Hall regime (strong
perpendicular magnetic field, low temperature) host states with highly
anisotropic transport properties. These occur when the filling factor $\nu
=2\pi n\ell ^{2}$ ($n$ is the electron density, $\ell =\sqrt{\hbar c/eB}$
the magnetic length and $B$ the magnetic field) is close to half integer
with numerator not too small. The strongest effects seem to occur for $\nu
\sim 9/2$, with similar phenomena present at $11/2,~13/2,...$ etc. Near
these filling factors a large asymmetry is observed in the diagonal
components of the resistivity tensor $\rho _{xx}$ and $\rho _{yy}$ that sets
in below approximately $100\;$mK in GaAs systems. The resistivity ratios $%
\rho _{xx}/\rho _{yy}$ may be as large as 3500\cite{lilly2}, although the
effect is exaggerated by system geometry\cite{simon}. The directions of
high/low resistance are clearly correlated with the GaAs crystal axes,
although the precise mechanism by which they are chosen is at present
unknown. For states near $\nu =11/2,~15/2,~19/2$, {\it etc}, the high/low
resistance directions may be rotated by application of an in-plane magnetic
field\cite{lilly2,pan}.

States leading to this anisotropic transport are likely to be related to 
{\it striped} states that were found in mean-field studies\cite
{shklovskii,moessner} of systems in which several Landau levels are filled,
and the highest occupied Landau level has a partial filling $\nu _{x}$ in
the range $0.35\lesssim \nu _{x}\lesssim 0.65$. Such ordering has been shown
to occur in exact diagonalization studies of finite size systems\cite{rezayi}%
. In a seminal theoretical work, the stability of this state to thermal and
quantum fluctuations was investigated by Fradkin and Kivelson\cite{fradkin},
who pointed out a powerful analogy between liquid crystals and quantum Hall
stripes. The analogy allows a classification of states according to
symmetries; these include stripe crystal, smectic, and nematic phases. As
will be shown below, at zero temperature mean-field theory predicts that the
stripe crystal is lowest in energy among these. However, it has been argued
that the smectic state may be stabilized by quantum\cite{fertig} or thermal 
\cite{macdonald} fluctuations, or both\cite{fradkin}. Finite temperature
studies\cite{fradkin2} of a model representing the nematic phase yield
impressive agreement with experiment of the resistance anisotropy at
temperatures that are not too low. Effects of in-plane magnetic fields\cite
{jungwirth,stanescu} have been studied and have provided some understanding
of the interchange of the high/low resistance directions, although the
different experimental behavior for $\nu =13/2,~17/2,...$ is still
unexplained.

Beyond transport studies, low-dimensional electron systems may be probed by
coupling to their collective modes, for example via inelastic light
scattering\cite{pinczuk} or surface acoustic waves\cite{willett}. These
collective modes for quantum Hall stripes are the subject of this study. Our
method will be the time-dependent Hartree-Fock approximation in the form
developed by C\^{o}t\'{e} and MacDonald \cite{cote}. The method requires a
static Hartree-Fock groundstate around which we can compute excitations. The
simplest form\cite{shklovskii} for such state is to treat the completely
filled $N-1$ Landau levels as inert, and form a one dimensional array of
alternating filled and empty guiding center states in the partially filled $%
N ${\it th} Landau level. In this approximation, the low energy Hamiltonian
for the partially filled level may be mapped to the lowest Landau level,
with a modified electron-electron interaction. This modification is
responsible for the low energy of stripe ordering in this system\cite
{shklovskii}.

We find, however, that uniform stripe states are unstable within the
Hartree-Fock approximation to formation of modulations along the stripes.
The resulting state is essentially an ordered array of one-dimensional
crystals, i.e., a ``stripe crystal'' \cite{fradkin}. Fig. 1 illustrates the
charge density for a stripe crystal phase. The amplitude of the density
modulation along the stripe is small; nevertheless, the energy gained in
going from uniform stripes to the stripe crystal is considerable. For
example, for $\nu _{x}=0.5$ in the $N=3$ Landau level, the striped phase is
found to have energy per particle of $-0.279691$ in units of $e^{2}/\kappa
\ell $ (here $\kappa $ is the dielectric constant; this will be our unit of
energy throughout this paper), while the stripe crystal has energy $%
-0.281465 $. For the parameters of Ref. \cite{lilly2}, the energy difference
between these two states is 112 mK, well above the temperatures for which
anisotropic transport is observed. Similar results are found at other values
of both $N$ and $\nu _{x}$.

The energy lowering in forming modulations along the stripes is largely an
intrastripe effect. For example, one may compute the energy of a stripe
crystal with a rectangular unit cell rather than the oblique one illustrated
in Fig. 1. The modulations of the stripes in this state are ``in-phase'',
requiring an additional Hartree energy. However, due to the weakness of the
modulations and the long-range nature of the Coulomb interaction, the
quantitative value of this energy cost is minuscule, of the order $%
10^{-8}e^{2}/\kappa \ell \sim 10^{-6}$ K. Thus, the chains may easily slide
past one another. Certainly, at any experimentally attainable temperature,
the crystal will melt into a series of thermally and quantum disordered
one-dimensional crystals\cite{fertig}.

In principle, at zero temperature Hartree-Fock theory predicts the system
locks into a stripe crystal. The collective modes around this state may be
characterized by wavevectors ${\bf k}=(k_{\parallel },k_{\perp })$, where $%
k_{\parallel }$ is the wavevector component parallel to the stripes and $%
k_{\perp }$ is the perpendicular component. The low energy collective modes
are phonons, and in principle are gapped everywhere except at ${\bf k}=0$.
In practice, because of the small energy scale associated with locking, we
find nearly gapless modes whenever $k_{\parallel }=0$, independent of $%
k_{\perp }$; the gaps are barely resolved by our numerical technique, and
are far below currently experimentally attainable temperature scales. The
low energy collective modes are thus highly reminiscent of what is expected
for a smectic state\cite{fradkin}.

Fig. 2 illustrates the phonon modes for several values of $k_{\perp }$ as a
function of $k_{\parallel }$, computed using the time-dependent Hartree-Fock
approximation (TDHFA) as described below. Several important features are
worth noting.

(1) The modes disperse linearly except for $k_{\perp }=0$, which disperses
more slowly. As shown below, this is consistent with a harmonic theory of a
charged smectic system in a magnetic field. The apparent absence of a gap
for $k_{\perp } \ne 0$ arises because in these collective modes the motion
of the electrons is parallel to the stripe direction. (This can be seen from
the eigenvector of the phonon mode from which one can compute the motion of
the stripes in real space. See Ref.\cite{movies} for details.) The gap is
then controlled by interactions of the modulations in different stripes,
which is very weak in this system.

(2) For larger values of $k_{\parallel }$, the modes become independent of $%
k_{\perp }$. Physically, this arises because the phonon modes are nearly
longitudinal for large $k_{\parallel }$, involving motion of the stripe
modulations but no significant motion of the positions of the stripes
relative to one another \cite{movies}. Since the stripe modulations
communicate so weakly, the relative phase of motion between stripes has
practically no effect on the energy of the mode -- hence, no $k_{\bot }$
dependence.

(3) As might be expected, a gap opens up near $k_{\parallel }=\pm \pi /b$,
where $b$ is the distance between modulations of a stripe. This leads to a
local maximum in the phonon dispersion. In light of (2) above, and as may be
seen explicitly in Fig. 2, the maximum is extremely flat along the $k_{\perp
}$ direction. As a result, there is a large phonon density of states (DOS)
at this energy, as illustrated in Fig. 3. Other minima and maxima appear in
the phonon dispersion which also contribute to structure in the DOS, most
notably a double peak at approximately half the energy of the $\pi /b$ peak.
Such structures may be observable in inelastic light scattering\cite{pinczuk}%
, and their detection would yield optical evidence of stripe ordering in
this system.

(4) A very low energy mode appears along the Brillouin zone boundary at $%
k_{\perp }=\pm \pi /a$ ($a$ is the separation between the stripes) near $%
k_{\parallel }=\pm \pi /2b.$ As $\left| \nu -0.5\right| $ increases, this
mode becomes soft (vanishing in energy) just above $\left| \nu -0.5\right|
=0.1$. This indicates a second order phase transition and increased
structure in the stripe state as one moves sufficiently away from
half-filling\cite{modecom}. This may indicate a second order phase
transition into a ``bubble phase'' \cite{shklovskii} or some precursor of
this phase. Alternatively, it may represent a buckling instability, in which
neighboring maxima within a stripe displace perpendicular to the stripe and
antiparallel to one another. (Such instabilities are known to occur at
Wigner crystal edges \cite{wcedge}.) The precise motion of the charge in the
soft mode is quite complex \cite{movies}. Work is currently underway to
determine the precise nature of the groundstate after the instability has
occured.

In addition to the low energy phonon modes, the stripe phases support
magnetoplasmon modes and spin wave modes, and we have explicitly computed
them in TDHFA. Fig. 4 illustrates an example of the magnetoplasmon modes
appearing as poles of the density response function in the first Brillouin
zone. The several apparent branches may be understood when the structure is
compared to analogous modes for a liquid state (no stripes) of the same
partially filled Landau level, illustrated by the solid lines in the same
figure. One may see that folding higher order Brillouin zones into the first
roughly generates the modes captured by the TDHFA. One may thus treat the
effect of stripe ordering on these high energy modes to a first
approximation as that of a periodic potential on an electron gas. A similar
effect occurs for the spin-wave modes. This does {\it not} occur for the
phonon modes because no such modes exist in the liquid state.

The presence of several branches of modes near small values of $k$ in
principle may be detected by optical or surface acoustic wave methods. Such
an observation would constitute a relatively direct demonstration of striped
ordering since it indicates zone-folding effects associated with a
unidirectional periodic modulation.

The remainder of this article is organized as follows. In Section \ref{hf},
the Hartree-Fock method used to generate mean-field states is briefly
discussed, and some more details of the results are provided. Section \ref
{tdhfa} briefly outlines the method used to obtain collective modes, and
presents the remainder of our results for collective modes, both in the
stripe state and, for comparison, in the liquid state. We conclude with a
summary (Section \ref{summary}). There are three Appendices. Appendix \ref
{app:tdhfa} provides some details of the proper formulation for TDHFA in
high Landau levels in general and striped states in particular. Analytic
expressions for collective modes of liquid states for partially filled
Landau levels are presented in Appendix \ref{app:liquid}. Appendix \ref
{harmonic} describes a simple elastic theory demonstrating that the results
of the TDHFA can be described at long wavelengths by a system with smectic
order.

\section{Hartree-Fock Approximation}

\label{hf}

In this Section we briefly review the Hartree-Fock approximation (HFA) as
developed in Ref. \cite{cote}; some further details are presented in
Appendix \ref{app:tdhfa}. The fundamental quantities in this approach are
the operators 
\begin{equation}
\rho _{n,m}^{\alpha ,\beta }\left( {\bf q}\right) ={\frac{1}{N_{\varphi }}}%
\sum_{X}\exp \left[ -iq_{x}X-iq_{x}q_{y}\ell ^{2}\right] c_{n,\alpha
,X}^{\dagger }c_{m,\beta ,X+q_{y}\ell ^{2}},  \label{rho}
\end{equation}
where $n,m$ denote Landau level indices, $N_{\varphi }$ is the Landau level
degeneracy, $X$ are guiding center coordinate quantum numbers and $\alpha
,\beta =\pm $ are spin indices. In the HFA, these quantities are evaluated
for a single Slater determinant state, which is accomplished by solving the
HFA equation of motion for the Green's function\cite{cote} 
\begin{equation}
G_{n,m}^{\alpha ,\beta }\left( {\bf G,}\tau \right) \equiv -{\frac{1}{%
N_{\varphi }}}\sum_{X}\left\langle Tc_{n,\alpha ,X}(\tau )c_{m,\beta
,X-G_{y}\ell ^{2}}^{\dag }\left( 0\right) \right\rangle \exp \left[
-iG_{x}X+iG_{x}G_{y}{\ell }^{2}/2\right] ,  \label{gf1}
\end{equation}
with $\left\{ {\bf G}\right\} $ the ensemble of reciprocal lattice vectors
of some assumed crystal structure. The HFA to the groundstate expectation
values of $\rho _{n,m}^{\alpha ,\beta }\left( {\bf q}\right) $ are non-zero
only for ${\bf q}$ on the reciprocal lattice, and are readily obtained from 
\cite{cote} 
\begin{equation}
\left\langle \rho _{n,m}^{\alpha ,\beta }\left( {\bf G}\right) \right\rangle
=G_{m,n}^{\beta ,\alpha }\left( {\bf G,}\tau =0^{-}\right) .
\end{equation}
Hartree-Fock energies, electron densities and response functions may be
computed from $\left\{ \left\langle \rho _{n,m}^{\alpha ,\beta }\left( {\bf G%
}\right) \right\rangle \right\} $.

For filling factors $\nu =2N+\widetilde{\nu }$, a further
simplification/approximation is to project the Hamiltonian into the single $%
N-th$ Landau level, which is formally appropriate when the electron-electron
interaction scale $e^{2}/\kappa \ell $ is much smaller than the cyclotron
energy $\omega _{c}$ (we take $\hslash =1$ throughout this paper). While in
experimental situations these energy scales are comparable, calculations\cite
{fertig2} retaining several Landau levels show that, for magnetic fields and
electron densities relevant to Ref. \cite{lilly}, Landau level mixing lowers
the Hartree-Fock energy by $\sim 10^{-4}e^{2}/\kappa \ell $ for the striped
state. This is sufficiently small to be neglected for our present study, and
we effectively retain only a single Landau level in our static HFA
calculations. Assuming also that there is no spin texture\cite{skyrmion} in
the ground-state, and denoting by the index $p$ the partially filled Landau
level, we have for $\widetilde{\nu }<1$

\begin{equation}
\left\{ 
\begin{array}{lcc}
\left\langle \rho _{m}^{\alpha }\left( {\bf G}\right) \right\rangle =\delta
_{{\bf G},0}, & {\rm if} & m<p; \\ 
\left\langle \rho _{m}^{+}\left( {\bf G}\right) \right\rangle \neq
0,\left\langle \rho _{m}^{+}\left( {\bf 0}\right) \right\rangle =\widetilde{%
\nu }, & {\rm if} & m=p; \\ 
\left\langle \rho _{m}^{-}\left( {\bf G}\right) \right\rangle =0, & {\rm if}
& m=p; \\ 
\left\langle \rho _{m}^{\alpha }\left( {\bf G}\right) \right\rangle =0, & 
{\rm if} & m>p,
\end{array}
\right.  \label{approx1}
\end{equation}
while for $\widetilde{\nu }>1,$ 
\begin{equation}
\left\{ 
\begin{array}{lcc}
\left\langle \rho _{m}^{\alpha }\left( {\bf G}\right) \right\rangle =\delta
_{{\bf G},0}, & {\rm if} & m<p; \\ 
\left\langle \rho _{m}^{+}\left( {\bf G}\right) \right\rangle =\delta _{{\bf %
G},0}, & {\rm if} & m=p; \\ 
\left\langle \rho _{m}^{-}\left( {\bf G}\right) \right\rangle \neq
0,\,\left\langle \rho _{m}^{-}\left( {\bf 0}\right) \right\rangle =%
\widetilde{\nu } & {\rm if} & m=p; \\ 
\left\langle \rho _{m}^{\alpha }\left( {\bf G}\right) \right\rangle =0, & 
{\rm if} & m>p.
\end{array}
\right.  \label{approx1b}
\end{equation}
We have defined $\rho _{n}^{\alpha }\left( {\bf G}\right) \equiv \rho
_{n,n}^{\alpha ,\alpha }\left( {\bf G}\right) $ to simplify the notation.

With our approximations, the filled levels are inert and cause only a shift
of the ground-state energy. Up to an unimportant constant, the interaction
energy per particle of the Hartree-Fock state is then 
\begin{equation}
E_{{\rm int}}^{HF}={\frac{1}{{2}{\nu _{x}}}}\sum_{{\bf G}}\left[
H_{pp}\left( {\bf G}\right) \left( 1-\delta _{{\bf G},0}\right)
-X_{pp}\left( {\bf G}\right) \right] \left| \left\langle \rho _{p}^{\alpha
}\left( {\bf G}\right) \right\rangle \right| ^{2},  \label{hfen}
\end{equation}
where $\alpha =+$ $\left( -\right) $ and $\nu _{x}=\widetilde{\nu }~(%
\widetilde{\nu }-1)$ if $\widetilde{\nu }<1$ ($\widetilde{\nu }>1$), and 
\begin{eqnarray}
H_{pp}\left( {\bf G}\right) &=&\left( \frac{e^{2}}{\kappa \ell }\right) 
\frac{1}{G\ell }e^{\frac{-G^{2}\ell ^{2}}{2}}\left[ L_{p}^{0}\left( \frac{%
G^{2}\ell ^{2}}{2}\right) \right] ^{2}, \\
X_{pp}\left( {\bf G}\right) &=&\left( \frac{e^{2}}{\kappa \ell }\right) 
\sqrt{2}\int_{0}^{\infty }dx\,e^{-x^{2}}\left[ L_{p}^{0}\left( x^{2}\right) %
\right] ^{2}J_{0}\left( \sqrt{2}xG\ell \right) ,  \nonumber
\end{eqnarray}
with $J_{0}\left( x\right) $ the Bessel function of order zero and $V\left( 
{\bf q}\right) =2\pi e^{2}/q$ the Fourier transform of the electron-electron
interaction, for which we use the unscreened Coulomb form. The functions $%
L_{n}^{m}\left( x\right) $ are generalized Laguerre polynomials.

To solve the Hartree-Fock equations, some guess is necessary for the crystal
structure of the groundstate to specify the set $\left\{ {\bf G}\right\} $.
The simplest structure for the stripes is a one-dimensional array with
lattice constant $a$. Writing $c_{p,\alpha ,X}\equiv c_{X}$, for $\alpha =+$
such states are characterized by order parameters 
\begin{equation}
\left\langle c_{X}^{\dag }c_{X^{\prime }}\right\rangle =\sum_{n=-\infty
}^{\infty }\Theta \left[ X-(n-\widetilde{\nu }/2)a\right] \Theta \left[ (n+%
\widetilde{\nu }/2)a-X\right] \delta _{X,X^{\prime }}.  \label{guiding}
\end{equation}
The density profile of the crystal phase is obtained from the relation 
\begin{equation}
\left\langle n\left( {\bf r}\right) \right\rangle =\frac{1}{2\pi \ell ^{2}}%
\sum_{{\bf G}}\left\langle \rho _{p}^{\alpha }\left( {\bf G}\right)
\right\rangle F_{p,p}\left( {\bf G}\right) e^{-i{\bf G}\cdot {\bf r}},
\label{patron}
\end{equation}
where $F_{p,p}\left( {\bf G}\right) $ is a form factor for electrons in
level $p$ (see Appendix A). One can also compute a ``density'' profile
corresponding to the guiding centers instead of the real density by using 
\begin{equation}
\left\langle n\left( {\bf r}\right) \right\rangle _{GC}=\sum_{{\bf G}%
}\left\langle \rho _{p}^{\alpha }\left( {\bf G}\right) \right\rangle e^{-i%
{\bf G}\cdot {\bf r}}.  \label{guidingc}
\end{equation}

Such states have been studied for a number of purposes\cite
{shklovskii,moessner,macdonald,jungwirth,stanescu} and provide a good first
approximation to the Hartree-Fock groundstate at the filling factors of
interest. However, within the HFA, this state is not stable and cannot be
used as a starting point for collective mode calculations: the resulting
response functions are unphysical. That this uniform stripe state is not a
minimum of the energy within the space of single Slater determinants may be
understood as follows. The interaction energy (Eq. (\ref{hfen})) for uniform
stripes may be written as $E_{{\rm int}}^{HF}={\frac{1}{2}}\sum_{X}^{\prime
}\varepsilon _{X}$, where $\varepsilon _{X}$ are the eigenvalues of the
Hartree-Fock Hamiltonian, and the prime indicates a sum over the $N_{p}$
lowest states, $N_{p}$ being the number of particles in the partially filled
level. The single-particle spectrum $\varepsilon _{X}$ has a well-defined
Fermi energy $E_{F}$ with eigenvalues arbitrarily close to it. By
introducing a one-dimensional modulation {\it along} the stripes, a gap is
opened at the Fermi energy, the eigenvalues $\varepsilon _{X}$ below $E_{F}$
are pushed down, and the total energy is lowered. The resulting state is an
array of one-dimensional crystals; i.e., a stripe crystal. The collective
modes presented below are all for such stripe crystal states.

We conclude this section with some remarks about the results of the HFA. An
example of the density modulation $\left\langle n\left( {\bf r}\right)
\right\rangle $ in a stripe crystal state is presented in Fig. 1. Results
for other Landau level indices $n\geq 2$ and partial fillings $\nu _{x}\sim
0.5$ are qualitatively similar to this. Two points are worth mentioning. (1)
The amplitude of the density modulations in real space are relatively weak,
across the stripes and even more so along them. Nevertheless, we will see in
the collective mode spectra clear signatures of both periodicities. It is
interesting to note that the weakness of the intrastripe modulations is due
mostly to the form factors of the $Nth$ Landau level; if one views the
``guiding center density'' as defined in Eq. (\ref{guidingc})  the
intrastripe modulations are quite pronounced (cf. \cite{movies}). (2) The
stripe crystal states studied here break particle-hole symmetry; there are
separate electron and hole stripe crystal solutions to the HFA which at $\nu
_{x}=1/2$ are degenerate. For $\nu _{x}<(>)~1/2$, the electron (hole)
crystal is lower in energy.

\section{Collective Modes in the TDHFA}

\label{tdhfa}

To obtain the dispersion relation of the collective excitations we compute
the matrix of response functions 
\begin{equation}
\chi _{n_{1},n_{2},n_{3},n_{4}}^{\alpha ,\beta ,\gamma ,\delta }\left( {\bf %
k+G},{\bf k}+{\bf G}^{\prime },\tau \right) =-N_{\varphi }\left\langle T%
\widetilde{\rho }_{n_{1},n_{2}}^{\alpha,\beta} \left( {\bf k}+{\bf G},\tau
\right) \widetilde{\rho }_{n_{3},n_{4}}^{\gamma,\delta} \left( -{\bf k}-{\bf %
G}^{\prime },0\right) \right\rangle ,  \label{chi}
\end{equation}
where $\widetilde{\rho }\left( {\bf q},\tau \right) =\rho \left( {\bf q}%
,\tau \right) -\left\langle \rho \left( {\bf q}\right) \right\rangle $ and $%
{\bf k}$ is a vector in the first Brillouin zone of the lattice. The
collective excitations appear as poles of the dynamical response functions
and their dispersion relation is obtained by tracking these poles for
several values of ${\bf k}$ in the first Brillouin zone. Since the order
parameters of Eq. (\ref{approx1}) were obtained in the Hartree-Fock
approximation, a conserving approximation for the response functions is
obtained in the time-dependent Hartree-Fock approximation (TDHFA). In Ref. 
\cite{cote}, it was shown that the equation of motion of this matrix of
response functions, in the TDHFA, can be written schematically as $\left[
I\left( \omega +i\delta \right) -A\right] \chi =B$ where $A$ and $B$ are
matrices that depend on matrix elements of the direct and exchange
interactions and on the order parameters $\left\{ \left\langle \rho
_{m}^{\alpha }\left( {\bf G}\right) \right\rangle \right\} $ only. All
response functions can then be obtained by solving numerically an eigenvalue
equation. In the simplest case (for the intra-Landau level excitation, for
example), $\chi $ consists of only one response function and accurate
results are easy to obtain. In other cases such as for the magnetoplasmon
excitations, response functions involving transitions to different Landau
levels are coupled and the matrix $\chi $ becomes rapidly very large. Our
method is thus limited by the size of the matrices $\chi $ that we can
handled numerically. Details of the calculation are given in Appendix \ref
{app:tdhfa}; here we present only the results. For concreteness, we focus on
a partially filled Landau level of index $N=3$ and spin $\alpha=+$, with $%
\widetilde{\nu} = 0.45.$ Results for other partial fillings, Landau level
indices, and spins are qualitatively similar.

To limit the size of the matrix $\chi $ we study the collective excitations
with $\omega (k=0)$ around $n\omega _{c}\pm mg^{\ast }\mu _{B}B$ with $m,n=0$
or $1$. We assume that $\omega _{c}$ is sufficiently large that coupling
among excitations near $n\omega _{c}$ and $n^{\prime }\omega _{c}$ may be
ignored if $n\neq n^{\prime }$. For comparison, we compute the same
dispersion relations (when they exist) in the liquid phase $i.e.$ in a
homogeneous phase with the same filling factor. The dispersions in that case
are simply obtained by replacing Eq.(\ref{approx1}) with $\left\langle \rho
_{3}^{+}\left( {\bf G=0}\right) \right\rangle =\widetilde{\nu }$ and setting
all other order parameters to zero\cite{com2}. Many, but not all, of the
results we find may be understood in terms liquid-like collective modes,
whose features have been folded into the first Brillouin zone by the
periodicity of the striped state.

There are five types of modes that we consider: (a) $n=0$: The phonon mode
(present in the stripe crystal phase only) appears as a pole of $\chi
_{nn}=\chi _{p,p,p,p}^{+,+,+,+}$ while the spin-wave mode $\omega
_{SW}\left( {\bf k}\right) $ is a pole of $\chi _{\sigma _{-}}=\chi
_{p,p,p,p}^{+,-,-,+}$, which, according to Larmor's theorem, should have $%
\omega _{SW}(0)=g^{\ast }\mu _{B}B.$ (b) $n=1$: There are three
magnetoplasmon modes in $\chi _{nn}$ that also appear in $\chi _{\sigma
_{z}}\equiv \left( \chi ^{++++}-\chi {^{++--}}-\chi ^{--++}+\chi
^{----}\right) /4$ but with different weight. These three magnetoplasmon
modes originate from the fact that there are three possible transitions with
pole around $\omega _{c}$, $i.e.$ $\left( 2,+\right) \rightarrow \left(
3,+\right) ,\left( 2,-\right) \rightarrow \left( 3,-\right) $ and $\left(
3,+\right) \rightarrow \left( 4,+\right) $. The Coulomb interaction mixes
these three modes, with the resulting dispersion branches being quite
complex even in the liquid phase. A spin-flip mode with $\delta S_{z}=+1$
appears as a pole of $\chi _{\sigma _{+}}\equiv \chi ^{-++-}$. The only
possible transition is $\left( 2,-\right) \rightarrow \left( 3,+\right) $
and there is correspondingly only a single branch in the dispersion. We will
refer to this mode as the $\omega _{SF+}$ mode. Finally, a pair of spin-flip
modes with $\delta S_{z}=-1$ appear as poles of $\chi _{\sigma _{-}}\equiv
\chi ^{+--+}$. These descend from transitions of the form $\left( 2,+\right)
\rightarrow \left( 3,-\right) ,\left( 3,+\right) \rightarrow \left(
4,-\right) $. We will refer to these two modes as the $\omega _{SF-}$ modes.

\subsection{Dispersion relation in the liquid phase}

Figs. 5 and 6 show the dispersion relation of the five modes for filling
factor $\widetilde{\nu }=0.45$ in the $N=3$ Landau level, in the liquid
phase. The complex dispersion relations are due in part to the generalized
Laguerre polynomial entering in the matrix elements of the Hartree-Fock
interaction which is responsible for the three minima appearing in all these
curves. In these figures (and all others that follow), we have substracted
the constant term $n\omega _{c}\pm mg^{\ast }\mu _{B}B$. Note that two of
the magnetoplasmon modes disperses from $\omega _{c}$ (as expected from
Kohn's theorem) while the third one is gapped. The higher energy mode that
disperses very rapidly is stronger in the density response function $\chi
_{nn}$ while the lowest-energy mode is stronger in $\chi _{\sigma _{z}}$.
The middle mode becomes very weak in both response functions as $%
k\rightarrow 0$.

>From Fig. 6, we see that the spin wave mode disperses from $g^{\ast }\mu
_{B}B$ as expected from Larmor's theorem. The inter-Landau-level excitations 
$\omega _{SF+}$ and $\omega _{SF-},$ however, have their gaps $\omega
_{SF+}\left( 0\right) =\omega _{c}-g^{\ast }\mu _{B}B$ and $\omega
_{SF-}\left( 0\right) =\omega _{c}+g^{\ast }\mu _{B}B$ strongly renormalized
by the self-energy and vertex corrections.

\subsection{Phonons}

For the stripe phase, we consider the configuration of Fig. 1 where the
electrons on one stripe are displaced with respect to the electrons on the
other stripes. This stripe crystal can be described by an oblique unit cell
with one electron or alternatively by a rectangular unit cell with two
electrons. In the inset of Fig. 2, we show the Brillouin zone of the oblique
unit cell that extend to $k_{\bot }\ell =\pm \,0.44$ and to $k_{\Vert }\ell
=\pm \,1.64$.

Unlike the homogeneous liquid phase, the stripe crystal phase can sustain a
phonon mode. The dispersion relation of this mode is presented in Fig. 2. As
discussed in the introduction, the most striking feature of the result is
the line of nearly gapless modes along $k_{\perp }=0$. Generically, for a
crystal one expects phonon modes to be gapped everywhere except at ${\bf k}%
=0 $. A careful examination of the small $k_{\Vert }$ limit is consistent
with this, although a precise determination of the gap is difficult because
the mode weights become very small in this limit. We estimate the gaps along
the $k_{\Vert }$ line to be in the range $10^{-7}-10^{-8}e^{2}/\kappa \ell $%
, which is far smaller than any experimentally accessible temperature.
Physically, this indicates the stripes are free to slide past one another
due to thermal fluctuations.

In the inset of Fig. 2, we show the dispersion relation along $k_{\Vert }$
for several values of $k_{\bot }$. One sees that the phonons disperse
linearly except for $k_{\bot }=0$ where they disperse more slowly. In
Appendix C, we show that this is consistent with a harmonic theory of a
charged smectic system in a magnetic field. Another point worth mentioning
is that for larger values of $k_{\parallel }$ the dispersion in $k_{\perp }$
becomes almost independent of $k_{\bot }$. By direct examination of the
charge motion in several such collective modes, we have found that this
arises because the phonon modes are nearly longitudinal for large $%
k_{\parallel }$; they do not involve significant motion of the positions of
the stripes relative to one another. Since the stripe modulations
communicate so weakly, the relative phase of motion between stripes has
practically no effect on the energy of the mode. As discussed in the
Introduction, this results in resonances in the collective mode density of
states that might be observed in inelastic light scattering. A particularly
strong such resonance occurs due to the additional flatness of the
dispersion near the Brillouin zone boundary along the direction of the
stripes (see Fig. 2) where a gap opens up separating the ``acoustic'' from
the ``optical'' modes.

Finally, as discussed in the Introduction, a soft mode appears that
indicates an instability of the modulated stripe state studied here for $%
\widetilde{\nu}$ just below $0.40$, suggesting at these lower fillings that
the correct groundstate will have more structure. We note that this
instability indicates a second-order transition, in contrast to the
first-order transition found near $\widetilde{\nu} \sim 0.36$ between stripe
and ``bubble'' states studied in Ref. \cite{shklovskii}. The result may
indicate that a precursor of the bubble phase develops within the stripe
phase, perhaps in which the bubbles are elongated rather than circular. It
is also possible that the stripes have a buckling instability in analogy
with similar behavior previously noted for Wigner crystal edges\cite{wcedge}.

\subsection{Higher Energy Modes}

Unlike the phonon mode, the four other excitations that we consider also
exist in the liquid phase. To understand the effect of the stripes, we plot
the dispersion relations obtained in the stripes and liquid phases together.
A few comments on these results are in order before we present them. For the
liquid, we fold the modes in the first Brillouin zone of the stripe crystal
and keep the lowest-energy branches $\left\{ \omega \left( {\bf k}+{\bf G}%
\right) \right\} $. Along the direction perpendicular to the stripes, the
lowest-energy branches correspond mostly to the functions $\omega \left(
k_{\bot }+nG_{\bot },k_{\Vert }=0\right) $ with $n=0,\pm 1,\pm 2,...$ In the
direction of the stripes, they correspond mostly to the curves $\omega
\left( k_{\bot }=nG_{\bot },k_{\Vert }\right) .$ In this case, however, the
curves with $n=\pm 1,\pm 2,\pm 3,...$ have the same energies in the liquid;
i.e., are degenerate. (The thick lines in the figures represent $n=0$ which
is not degenerate). This degeneracy is sometimes lifted in the stripe phase.
Note that this Brillouin zone folding of the liquid dispersions introduces a
large number of branches. It is not possible to track all the corresponding
poles in the stripe phase. We thus sometimes show only a small subset of
these modes corresponding to low-energy excitations. Because we keep only
the most intense poles and because the relative intensity changes as ${\bf k}
$ spans the Brillouin zone, the dispersions sometimes appear discontinuous
for the stripe phase; this is because the mode weights fall below our
threshold for plotting them. Note that the zone-folding effects lead to the
presence of several branches near small values of ${\bf k.}$ In principle,
these excitations could be detected by optical or surface acoustic wave
methods and would thus represent a direct demonstration of the stripe
ordering.

\subsubsection{Spin Waves}

For GaAs systems, under most circumstances $g^{\ast }\mu _{B}B<<\hbar \omega
_{c}$, so that spin waves are the lowest energy modes after the phonons.
Figs. 7 and 8 show the dispersion relation obtained for the spin waves. The
most striking difference between the zone-folded liquid results and the spin
waves of the stripe state is a dramatic anisotropy in the gap opening at the
Brillouin zone boundary. This gap is much larger at the boundary for large $%
k_{\perp }$ than the corresponding one for large $k_{\parallel }$.
Certainly, part of the explanation is that the density modulations
responsible for the latter is much smaller than that of the former. However,
the gap at large $k_{\perp }$ is much larger than, for example, the
corresponding gap in the magnetoplasmons, discussed below. This strong
many-body effect may be related to electrons at the stripe edges having a
small local spin stiffness relative to those in the liquid state or in the
center of the stripes. In any case, this many-body effect results in a
branch of spin waves with surprisingly low energy.

\subsubsection{Magnetoplasmons}

Figs. 4 and 9 show the dispersion relation obtained in the stripe phase for
the magnetoplasmon modes. To capture all the three branches, we show poles
obtained from both the density and spin response functions $\chi _{nn}$ and $%
\chi _{\sigma _{z}}$. Because the three corresponding branches in the liquid
are almost flat at large wave vector, the folding of the modes in the first
Brillouin zone introduces many branches at small energy. It is quite clear,
however, that the dispersion obtained in the stripe phase follows closely
that of the liquid, with small gaps at the Brillouin zone edges and some
lifting of degeneracy in the $k_{\parallel }$ direction.

\subsubsection{Spin Flip Excitations}

Some of the spin flip excitations seem to follow behavior reminiscent of our
results for the magnetoplasmons, closely following the liquid results,
whereas others undergo strong many-body renormalizations, as we found for
spin waves. Fig. 10 shows the dispersion for the $\omega _{SF-}$ mode. (In
these figures, we show only the low-energy excitations because the liquid
modes become very complicated at higher energies.) One can see the direct
correpondance between the liquid phase dispersions and those of the stripe
states. As for the magnetoplasmons, these differ by gap openings and the
lifting of degeneracies. For the $\omega _{SF+}$ mode, as for the spin wave
mode, the dispersion relation of the lowest-energy branches are quite close
to the corresponding liquid result. For higher branches, however, the
stripes ordering lead to important changes as can be seen in Fig. 11. The
difference between the behaviors of $\omega _{SF+}$ and $\omega _{SF-}$ is
very likely related to the presence of two branches of the former in the
liquid state which may be mixed by the stripe ordering, whereas only a
single branch exists in the latter.

\section{Summary}

\label{summary}

In this work we have studied collective mode of stripe states for quantum
Hall systems. The lowest energy states are phonons, with a line in the
Brillouin zone of extremely low energy states, making the resulting
low-energy physics of this system that of a charged, two-dimensional smectic
in a magnetic field. We also found signatures in the phonon density of
states indicative of stripe ordering that should be detectable in inelastic
light scattering, and a soft mode that indicates an instability of the
stripe state for partial fillings sufficiently far from $1/2$. Results for
spin waves, magnetoplasmon, and spin-flip excitations were also presented,
which in a first approximation could be understood in terms of zone-folding
of corresponding excitations for the liquid state. Some of these, however,
underwent strong renormalizations due to electron-electron interactions; in
particular, we found a surprisingly low-energy branch in the spin wave
spectrum due to this effect.

The form of the low energy physics of this system has important consequences
for quantum fluctuation effects on the stripe crystal state, particularly
the stability of the crystal as well as pinning by disorder. Some of this
has been discussed previously\cite{fertig}; a more detailed study will be
presented in future work.

{\it Acknowledgements} -- The authors would like to thank Dr. Hangmo Yi for
many useful discussions, as well as helpful discussions with Allan MacDonald
and Matthew Fisher in the initial stages of this work. This work was
supported by NSF Grant Nos. DMR-98-70681 and PHY94-07194, by the Research
Corporation and by the National Science and Engineering Research Council of
Canada. HAF thanks the Institute for Theoretical Physics at Santa Barbara,
where this work was initiated, for its hospitality.

\appendix

\section{Details of the TDHFA}

\label{app:tdhfa}

In this Appendix we discuss the proper formulation of the TDHFA in high
Landau levels. The basic approach follows that of Ref. \cite{cote}; however,
there are important details involved in computing inter-Landau level
excitations that have been treated incorrectly\cite{cote,kallin} in the
literature\cite{brey}, leading to results that do not correctly include the
exchange self-energy corrections to these excitations. We present here a
correct formulation of the TDHFA that avoids such errors and respects Kohn's
theorem\cite{kohn}.

\subsection{Static Hartree-Fock Approximation}

We begin by briefly reviewing the relevant equations for HFA that will be
needed in our formulation of the TDHFA; details may be found in Ref. \cite
{cote}. Our model HF Hamiltonian is 
\begin{equation}
H_{HF}=N_{\varphi }\sum_{n,\alpha }\varepsilon _{n,\alpha }\rho _{n,\alpha
}\left( 0\right) +N_{\varphi }\sum_{n,\alpha }\sum_{{\bf G}}U_{n}^{\alpha
}\left( {\bf G}\right) \rho _{n}^{\alpha }\left( {\bf G}\right) ,
\label{hamiltonian}
\end{equation}
where 
\begin{equation}
\varepsilon _{n,\alpha }=\left( n+1/2\right) \omega _{c}-\alpha g^{\ast }\mu
_{B}B/2.
\end{equation}
The Hartree-Fock effective potential $U_{n}^{\alpha }\left( {\bf G}\right) $
is given by 
\begin{eqnarray}
U_{n}^{\alpha }\left( {\bf G}\right) &=&\sum_{m}\sum_{\beta }\left[ H\left(
m,m,n,n;{\bf G}\right) -\delta _{\alpha ,\beta }X\left( m,n,n,m;{\bf G}%
\right) \right] \left\langle \rho _{m}^{\beta }\left( -{\bf G}\right)
\right\rangle  \label{effectivepot} \\
&\equiv &\sum_{m}\sum_{\beta }\left[ H_{m,n}\left( {\bf G}\right) -\delta
_{\alpha ,\beta }X_{m,n}\left( {\bf G}\right) \right] \left\langle \rho
_{m}^{\beta }\left( -{\bf G}\right) \right\rangle .  \nonumber
\end{eqnarray}
For completeness, we give here the form of the Hartree and Fock interactions
that enter into the calculation of the self-energy corrections to the
collective excitations: 
\begin{eqnarray}
H_{m,n}\left( {\bf q}\right) &=&\left( \frac{e^{2}}{\kappa \ell }\right) 
\frac{1}{q\ell }e^{\frac{-q^{2}\ell ^{2}}{2}}L_{m}^{0}\left( \frac{q^{2}\ell
^{2}}{2}\right) L_{n}^{0}\left( \frac{q^{2}\ell ^{2}}{2}\right) , \\
X_{m,n}\left( {\bf q}\right) &=&\left( \frac{e^{2}}{\kappa \ell }\right) 
\sqrt{2}\left( \frac{n!}{m!}\right) \int_{0}^{\infty }dx\,x^{2\left(
m-n\right) }e^{-x^{2}}\left[ L_{n}^{m-n}\left( x^{2}\right) \right]
^{2}J_{0}\left( \sqrt{2}xq\ell \right) \text{ (for }n\leq m\text{)}. 
\nonumber
\end{eqnarray}
For $n>m,$ we use $X_{n,m}\left( {\bf q}\right) =X_{m,n}\left( {\bf q}%
\right) $.

The effective interactions appearing in Eq.(\ref{effectivepot}) are a subset
of the more general form 
\begin{eqnarray}
H\left( n_{1},n_{2},n_{3},n_{4};{\bf q}\right) &=&\left( \frac{e^{2}}{\kappa
\ell }\right) \left( \frac{1}{2\pi e^{2}\ell }\right) V\left( {\bf q}\right)
F_{n_{1},n_{2}}\left( {\bf q}\right) F_{n_{3},n_{4}}\left( -{\bf q}\right) ,
\label{hartree} \\
X\left( n_{1},n_{2},n_{3},n_{4};{\bf q}\right) &=&\left( \frac{e^{2}}{\kappa
\ell }\right) \left( \frac{\ell }{e^{2}}\right) \int \frac{d^{2}q^{\prime }}{%
\left( 2\pi \right) ^{2}}V\left( {\bf q}^{\prime }\right)
F_{n_{1},n_{2}}\left( {\bf q}^{\prime }\right) F_{n_{3},n_{4}}\left( -{\bf q}%
^{\prime }\right) e^{-i{\bf q}\times {\bf q}^{\prime }\ell ^{2}},
\label{fock}
\end{eqnarray}
that we need to derive the TDHFA. We take $V\left( {\bf q}\right) =2\pi
e^{2}/q$. (We use the two-dimensional cross product as a short form for $%
{\bf q}\times {\bf G\equiv }q_{x}G_{y}-q_{y}G_{x}$). These interactions
contain the form factors 
\begin{equation}
F_{n,m}\left( {\bf q}\right) =\left( \frac{m!}{n!}\right) ^{1/2}\left( \frac{%
\left( -q_{y}+iq_{x}\ell \right) }{\sqrt{2}}\right) ^{n-m}\exp \left[ \frac{%
-q^{2}\ell ^{2}}{4}\right] L_{m}^{n-m}\left( \frac{q^{2}\ell ^{2}}{2}\right)
,  \label{fnmq}
\end{equation}
for $m\leq n$, where $L_{n}^{\alpha }\left( x\right) $ is the generalized
Laguerre polynomial. Note that $F_{n,m}\left( {\bf q}\right) =\left[
F_{m,n}\left( -{\bf q}\right) \right] ^{\ast }.$ We remark that the
effective potential in any Landau level $n,\alpha $ depends on the
occupation of the other levels, as does the energy of the electrons in that
level. This self-energy shift differs from one level to another, and makes
an important contribution to the energy of inter-Landau level excitations.

The single particle Green's function of Eq.(\ref{gf1}) obeys, under the
Hamiltonian of Eq. (\ref{hamiltonian}), the equation of motion 
\begin{equation}
\left[ i\omega _{n}-\left( \varepsilon _{n}^{\alpha }-\mu \right) \right]
G_{n}^{\alpha }\left( {\bf G,}i\omega _{n}\right) -\sum_{{\bf G}^{\prime
}}W_{n}^{\alpha }\left( {\bf G}-{\bf G}^{\prime }\right) G_{n}^{\alpha
}\left( {\bf G}^{\prime }{\bf ,}i\omega _{n}\right) =\delta _{{\bf G},0},
\label{motion}
\end{equation}
where $\mu $ is the chemical potential and 
\begin{equation}
W_{n}^{\alpha }\left( {\bf G}-{\bf G}^{\prime }\right) \equiv U_{n}^{\alpha
}\left( {\bf G}^{\prime }-{\bf G}\right) e^{i{\bf G}\times {\bf G}^{\prime
}\ell ^{2}/2}.
\end{equation}
Eq.(\ref{motion}) can be solved numerically to compute the densities $%
\left\langle \rho _{n}^{\sigma }\left( {\bf G}\right) \right\rangle $ as
explained in Ref. \cite{cote}.

\subsection{Time-Dependent Hartree-Fock Approximation}

The two-particle Green's functions are defined by Eq. (\ref{chi}). In the
TDHFA, they obey an equation of motion that we write as\cite{cote} 
\begin{align}
& \left[ i\Omega _{n}+\left( \varepsilon _{n_{1},\alpha }-\varepsilon
_{n_{2},\beta }\right) \right] \chi _{n_{1},n_{2},n_{3},n_{4}}^{(0)\alpha
,\beta ,\gamma ,\delta }\left( {\bf k+G,k+G}^{\prime },\Omega _{n}\right)
\label{chi00} \\
& +\sum_{{\bf G}^{\prime \prime }}\left[ \gamma _{{\bf G},{\bf G}^{\prime
\prime }}^{\ast }\left( {\bf k}\right) U_{n_{1}}^{\alpha }\left( {\bf G}%
^{\prime \prime }-{\bf G}\right) -\gamma _{{\bf G},{\bf G}^{\prime \prime
}}\left( {\bf k}\right) U_{n_{2}}^{\beta }\left( {\bf G}^{\prime \prime }-%
{\bf G}\right) \right] \chi _{n_{1},n_{2},n_{3},n_{4}}^{(0)\alpha ,\beta
,\gamma ,\delta }\left( {\bf k+G}^{\prime \prime }{\bf ,k+G}^{\prime
},\Omega _{n}\right)  \nonumber \\
& =\delta _{n_{1},n_{4}}\delta _{n_{2},n_{3}}\delta _{\alpha ,\delta }\delta
_{\beta ,\gamma }\left[ \gamma _{{\bf G},{\bf G}^{\prime }}^{\ast }\left( 
{\bf k}\right) \left\langle \rho _{n_{1}}^{\alpha }\left( {\bf G-G}^{\prime
}\right) \right\rangle -\gamma _{{\bf G},{\bf G}^{\prime }}\left( {\bf k}%
\right) \left\langle \rho _{n_{2}}^{\beta }\left( {\bf G-G}^{\prime }\right)
\right\rangle \right] ,  \nonumber
\end{align}
\begin{align}
\widetilde{\chi }_{n_{1},n_{2},n_{3},n_{4}}^{\alpha ,\beta ,\gamma ,\delta
}\left( {\bf k+G,k+G}^{\prime },\Omega _{n}\right) & =\chi
_{n_{1},n_{2},n_{3},n_{4}}^{(0)\alpha ,\beta ,\gamma ,\delta }\left( {\bf %
k+G,k+G}^{\prime },\Omega _{n}\right)  \label{chitilde00} \\
& -\sum_{n_{5},\cdots n_{8}}\sum_{\eta \nu }\sum_{{\bf G}^{\prime \prime
}}\chi _{n_{1},n_{2},n_{5},n_{6}}^{(0)\alpha ,\beta ,\eta ,\nu }\left( {\bf %
k+G,k+G}^{\prime \prime },\Omega _{n}\right)  \nonumber \\
& \times X\left( n_{7},n_{6},n_{5},n_{8};{\bf k}+{\bf G}^{\prime \prime
}\right) \widetilde{\chi }_{n_{7},n_{8},n_{3},n_{4}}^{\nu ,\eta ,\gamma
,\delta }\left( {\bf k+G}^{\prime \prime }{\bf ,k+G}^{\prime },\Omega
_{n}\right) ,  \nonumber
\end{align}
and 
\begin{align}
\chi _{n_{1},n_{2},n_{3},n_{4}}^{\alpha ,\beta ,\gamma ,\delta }\left( {\bf %
k+G,k+G}^{\prime },\Omega _{n}\right) & =\widetilde{\chi }%
_{n_{1},n_{2},n_{3},n_{4}}^{\alpha ,\beta ,\gamma ,\delta }\left( {\bf %
k+G,k+G}^{\prime },\Omega _{n}\right)  \label{chi000} \\
& +\sum_{n_{5},\cdots n_{8}}\sum_{\eta \nu }\sum_{{\bf G}^{\prime \prime }}%
\widetilde{\chi }_{n_{1},n_{2},n_{5},n_{6}}^{\alpha ,\beta ,\nu ,\nu }\left( 
{\bf k+G,k+G}^{\prime \prime },\Omega _{n}\right)  \nonumber \\
& \times H\left( n_{5},n_{6},n_{7},n_{8};{\bf k}+{\bf G}^{\prime \prime
}\right) \chi _{n_{7},n_{8},n_{3},n_{4}}^{\eta ,\eta ,\gamma ,\delta }\left( 
{\bf k+G}^{\prime \prime }{\bf ,k+G}^{\prime },\Omega _{n}\right) , 
\nonumber
\end{align}
where $\Omega _{n}$ is a Boson Matsubara frequency and 
\begin{equation}
\gamma _{{\bf G},{\bf G}^{\prime }}\left( {\bf k}\right) \equiv e^{i\left( 
{\bf k}+{\bf G}\right) \times \left( {\bf k}+{\bf G}^{\prime }\right) \ell
^{2}/2}.  \label{gamma}
\end{equation}
Eqs.(\ref{chi00} - \ref{chi000}) are equivalent to the result of summing
ladder and bubble diagrams in a perturbative expansion of $\chi $. Note that
the only information required in these equations is the groundstate density $%
<\rho _{n}^{\alpha }({\bf G})>$. The equations couple together an infinite
set of response functions; as discussed in Ref. \cite{cote}, when truncated
appropriately they may be cast in a matrix form for numerical solution. In
the next few sections, we describe truncations and simplifications that are
appropriate for computing various collective modes.

\subsection{Equation of motion for $\protect\chi ^{(0)}$}

It follows from Eq.(\ref{chi00}) that the only non zero $\chi ^{(0)}$ must
be of the form $\chi _{n,m,m,n}^{(0),\alpha ,\beta ,\beta ,\alpha }$.
Written in matrix notation (with the reciprocal lattice vectors ${\bf G},%
{\bf G}^{\prime }$ being the matrix indices), the equation of motion for $%
\chi ^{(0)}$ is then 
\begin{equation}
\left[ i\Omega _{n}I-F_{n,m}^{\alpha ,\beta }\left( {\bf k}\right) \right]
\chi _{n,m,m,n}^{(0),\alpha ,\beta ,\beta ,\alpha }\left( {\bf k},\omega
\right) =B_{n,m}^{\alpha ,\beta }\left( {\bf k}\right) ,  \label{eqchizero}
\end{equation}
where 
\begin{equation}
\left[ F_{n,m}^{\alpha ,\beta }\left( {\bf k}\right) \right] _{{\bf G},{\bf G%
}^{\prime }}\equiv \left( \varepsilon _{m,\beta }-\varepsilon _{n,\alpha
}\right) \delta _{{\bf G},{\bf G}^{\prime }}-U_{n}^{\alpha }\left( {\bf G}%
^{\prime }-{\bf G}\right) \gamma _{{\bf G},{\bf G}^{\prime }}^{\ast }\left( 
{\bf k}\right) +U_{m}^{\beta }\left( {\bf G}^{\prime }-{\bf G}\right) \gamma
_{{\bf G},{\bf G}^{\prime }}\left( {\bf k}\right)
\end{equation}
and 
\begin{equation}
\left[ B_{n,m}^{\alpha ,\beta }\left( {\bf k}\right) \right] _{{\bf G},{\bf G%
}^{\prime }}\equiv \gamma _{{\bf G},{\bf G}^{\prime }}^{\ast }\left( {\bf k}%
\right) \left\langle \rho _{n}^{\alpha }\left( {\bf G-G}^{\prime }\right)
\right\rangle -\gamma _{{\bf G},{\bf G}^{\prime }}\left( {\bf k}\right)
\left\langle \rho _{m}^{\beta }\left( {\bf G-G}^{\prime }\right)
\right\rangle .  \label{bnm}
\end{equation}
The size of these matrices depends on the number of reciprocal lattice
vectors that are kept in the numerical calculation.

\subsection{Equation of motion for $\widetilde{\protect\chi }$}

Since $\chi ^{\left( 0\right) }$ is of the form $\chi _{n,m,m,n}^{(0),\alpha
,\beta ,\beta ,\alpha },$ Eq.(\ref{chitilde00}) for $\widetilde{\chi }$ can
be simplified to 
\begin{eqnarray}
\widetilde{\chi }_{n_{1},n_{2},n_{3},n_{4}}^{\alpha ,\beta ,\gamma ,\delta
}\left( {\bf k},\omega \right) &=&\chi
_{n_{1},n_{2},n_{2},n_{1}}^{(0),\alpha ,\beta ,\beta ,\alpha }\left( {\bf k}%
,\omega \right) \delta _{n_{1},n_{4}}\delta _{n_{3},n_{2}}\delta _{\alpha
,\delta }\delta _{\beta ,\gamma }  \label{chitilde1} \\
&&-\chi _{n_{1},n_{2},n_{2},n_{1}}^{(0),\alpha ,\beta ,\beta ,\alpha }\left( 
{\bf k},\omega \right) \sum_{n_{5},n_{6}}\left[ X_{n_{5},n_{1},n_{2},n_{6}}%
\left( {\bf k}\right) \widetilde{\chi }_{n_{5},n_{6},n_{3},n_{4}}^{\alpha
,\beta ,\gamma ,\delta }\left( {\bf k},\omega \right) \right] ,  \nonumber
\end{eqnarray}
where the matrix 
\begin{equation}
\left[ X_{_{n_{1},n_{2},n_{3},n_{4}}}\left( {\bf k}\right) \right] _{{\bf G},%
{\bf G}^{\prime }}\equiv X\left( n_{1},n_{2},n_{3},n_{4};{\bf k+G}\right)
\delta _{{\bf G},{\bf G}^{\prime }}.
\end{equation}

>From Eq.(\ref{chitilde1}), it is clear that, in the spin indices, $%
\widetilde{\chi }$ must be of the form $\widetilde{\chi }^{\alpha ,\beta
,\beta ,\alpha }$ and that $\widetilde{\chi }_{n_{1},n_{2},n_{3},n_{4}}^{%
\alpha ,\beta ,\gamma ,\delta }\neq 0,\,$only if $\chi
_{n_{1},n_{2},n_{2},n_{1}}^{(0)\text{ }\alpha ,\beta ,\beta ,\alpha }\neq 0$%
. Since we are working in the strong magnetic field limit ($\omega
_{c}>>e^{2}/\kappa \ell $), we will assume that a response function with
poles around $n\omega _{c}$ is only coupled to other response functions
poles near the same frequency \cite{kallin}. {\it Thus, we truncate our
equations by including coupling among response functions of the form $%
\widetilde{\chi }_{n_{1}+m,n_{2}+m,n_{3},n_{4}}. $ for different values of $%
m $}. We remark here that coupling {\it all} response functions with pole
around the same frequency is {\it essential} to recover Kohn's theorem for
the cyclotron mode.

Eq.(\ref{chitilde1}) is now simplified to 
\begin{align}
\widetilde{\chi }_{n_{1}+m,n_{2}+m,n_{2},n_{1}}^{\alpha ,\beta ,\beta
,\alpha }\left( {\bf k},\omega \right) & =\chi
_{n_{1},n_{2},n_{2},n_{1}}^{(0),\alpha ,\beta ,\beta ,\alpha }\left( {\bf k}%
,\omega \right) \delta _{m,0}-\chi
_{n_{1}+m,n_{2}+m,n_{2}+m,n_{1}+m}^{(0),\alpha ,\beta ,\beta ,\alpha }\left( 
{\bf k},\omega \right)  \label{chitilde2} \\
& \times \sum_{m^{\prime }}\left[ X_{n_{1}+m^{\prime
},n_{1}+m,n_{2}+m,n_{2}+m^{\prime }}\left( {\bf k}\right) \widetilde{\chi }%
_{n_{1}+m^{\prime },n_{2}+m^{\prime },n_{2},n_{1}}^{\alpha ,\beta ,\beta
,\alpha }\left( {\bf k},\omega \right) \right] .  \nonumber
\end{align}
Because we consider only the special case where all Landau levels below $p$
are completely filled and $p$ is partially filled, Eq.(\ref{bnm}) implies
that 
\begin{equation}
\left\{ 
\begin{array}{lll}
\chi _{p+m,p+m,p+m,p+m}^{(0),\alpha ,\beta ,\beta ,\alpha }\neq 0 & \text{%
only if} & m=0, \\ 
\chi _{p+m,p+m+1,p+m+1,p+m}^{(0),\alpha ,\beta ,\beta ,\alpha }\neq 0 & 
\text{only if} & m=-1,0, \\ 
\chi _{p+m,p+m+2,p+m+2,p+m}^{(0),\alpha ,\beta ,\beta ,\alpha }\neq 0 & 
\text{only if} & m=-2,-1,0,
\end{array}
\right.  \label{approx5}
\end{equation}
and so one. $\widetilde{\chi }$ will thus be coupled to one, two, three or
more other $\widetilde{\chi }^{\prime }s$ depending on the value of $m$ and
also on the number of levels filled below $p$. For example, we only need to
consider the response function $\widetilde{\chi }_{p,p,p,p}^{\alpha ,\beta
,\beta ,\alpha }$ for the intra-Landau-level excitation. Its equation of
motion is thus simply 
\begin{equation}
\widetilde{\chi }_{p,p,p,p}^{\alpha ,\beta ,\beta ,\alpha }\left( {\bf k}%
,\omega \right) =\chi _{p,p,p,p}^{(0),\alpha ,\beta ,\beta ,\alpha }\left( 
{\bf k},\omega \right) -\chi _{p,p,p,p}^{(0),\alpha ,\beta ,\beta ,\alpha
}\left( {\bf k},\omega \right) X_{p}^{(0)}\left( {\bf k}\right) \text{ }%
\widetilde{\chi }_{p,p,p,p}^{\alpha ,\beta ,\beta ,\alpha }\left( {\bf k}%
,\omega \right) ,  \label{chitilde3}
\end{equation}
where we have defined the diagonal matrix 
\begin{equation}
\left[ X_{n}^{(0)}\left( {\bf k}\right) \right] _{{\bf G},{\bf G}^{\prime
}}\equiv X\left( n,n,n,n;{\bf k+G}\right) \delta _{{\bf G},{\bf G}^{\prime
}}.
\end{equation}
With Eq.(\ref{eqchizero}), Eq.(\ref{chitilde3}) becomes 
\begin{equation}
\left[ i\Omega _{n}I-F_{p,p}^{\alpha ,\beta }\left( {\bf k}\right)
+B_{p,p}^{\alpha ,\beta }\left( {\bf k}\right) X_{p}^{(0)}\left( {\bf k}%
\right) \right] \widetilde{\chi }_{p,p,p,p}^{\alpha ,\beta ,\beta ,\alpha
}\left( {\bf k},\omega \right) =B_{p,p}^{\alpha ,\beta }\left( {\bf k}%
\right) .  \label{chitilde6}
\end{equation}

For $m=1$ (inter-Landau-level excitations), there are only four non zero $%
\widetilde{\chi }$ with poles around $+\omega _{c}.$ To deal with this case,
it is helpful to define the block matrices (which we denote by the symbol $%
\overline{\widetilde{\chi }}$ to distinguish it from the simple matrix $%
\widetilde{\chi }$) 
\begin{equation}
\overline{\widetilde{\chi }}\equiv \left[ 
\begin{array}{cc}
\widetilde{\chi }_{p,p+1,p+1,p}^{\alpha ,\beta ,\beta ,\alpha } & \widetilde{%
\chi }_{p,p+1,p,p-1}^{\alpha ,\beta ,\beta ,\alpha } \\ 
\widetilde{\chi }_{p-1,p,p+1,p}^{\alpha ,\beta ,\beta ,\alpha } & \widetilde{%
\chi }_{p-1,p,p,p-1}^{\alpha ,\beta ,\beta ,\alpha }
\end{array}
\right] ,  \label{chinonzero}
\end{equation}
and 
\begin{equation}
\overline{I}\equiv \left[ 
\begin{array}{cc}
I & 0 \\ 
0 & I
\end{array}
\right] .
\end{equation}
In terms of these matrices, Eq.(\ref{chitilde00}) simplifies to

\begin{equation}
\left[ i\Omega _{n}\overline{I}-\overline{F}^{\alpha ,\beta }\left( {\bf k}%
\right) +\overline{B}^{\alpha ,\beta }\left( {\bf k}\right) \overline{X_{p}}%
\left( {\bf k}\right) \right] \overline{\widetilde{\chi }}^{\alpha ,\beta
,\beta ,\alpha }\left( {\bf k},\omega \right) =\overline{B}^{\alpha ,\beta
}\left( {\bf k}\right) ,  \label{spinflip}
\end{equation}
where 
\begin{equation}
\overline{F}^{\alpha ,\beta }\left( {\bf k}\right) \equiv \left[ 
\begin{array}{cc}
F_{p,p+1}^{\alpha ,\beta }\left( {\bf k}\right) & 0 \\ 
0 & F_{p-1,p}^{\alpha ,\beta }\left( {\bf k}\right)
\end{array}
\right] ,
\end{equation}
\begin{equation}
\overline{B}^{\alpha ,\beta }\left( {\bf k}\right) \equiv \left[ 
\begin{array}{cc}
B_{p,p+1}^{\alpha ,\beta }\left( {\bf k}\right) & 0 \\ 
0 & B_{p-1,p}^{\alpha ,\beta }\left( {\bf k}\right)
\end{array}
\right] ,
\end{equation}
and 
\begin{equation}
\overline{X}_{n}\equiv \left[ 
\begin{array}{cc}
X_{n}^{(1)} & X_{n}^{(4)} \\ 
X_{n}^{(2)} & X_{n}^{(3)}
\end{array}
\right] ,
\end{equation}
with 
\begin{eqnarray}
\left[ X_{n}^{(1)}\left( {\bf k}\right) \right] _{{\bf G},{\bf G}^{\prime }}
&\equiv &X\left( n,n,n+1,n+1;{\bf k+G}\right) \delta _{{\bf G},{\bf G}%
^{\prime }}; \\
\left[ X_{n}^{(2)}\left( {\bf k}\right) \right] _{{\bf G},{\bf G}^{\prime }}
&\equiv &X\left( n,n-1,n,n+1;{\bf k+G}\right) \delta _{{\bf G},{\bf G}%
^{\prime }}; \\
\left[ X_{n}^{(3)}\left( {\bf k}\right) \right] _{{\bf G},{\bf G}^{\prime }}
&\equiv &X\left( n-1,n-1,n,n;{\bf k+G}\right) \delta _{{\bf G},{\bf G}%
^{\prime }}; \\
\left[ X_{n}^{(4)}\left( {\bf k}\right) \right] _{{\bf G},{\bf G}^{\prime }}
&\equiv &X\left( n-1,n,n+1,n;{\bf k+G}\right) \delta _{{\bf G},{\bf G}%
^{\prime }}.
\end{eqnarray}
The solutions to Eq.(\ref{spinflip}) can be used to compute both density and
spin-flip response functions (e.g., $\chi _{p,p+1,p+1,p}^{+--+}$).

In principle, we can deal in the same manner with excitations around $%
2\omega _{c}.$ These would involve solving a $3\times 3$ block matrix in $%
\widetilde{\chi }$ (depending upon the number of filled levels below $p$).
Since each block in these matrices is itself a matrix whose size depends on
the number of reciprocal lattice vectors that we keep in the calculation,
solving for higher-energy excitations becomes difficult numerically. We will
thus be satisfied here with the solution for intra and inter-Landau-level
response functions with poles around zero or $\omega _{c}$ (shifted, of
course, by the Zeeman energy if spin flip excitations are considered). This
includes the important case of phonons and spin-wave excitations in the
partially filled level, as well as the cyclotron modes around $\omega _{c}$
and spin-flip modes from or to the partially filled Landau level.

\subsection{Equation of motion for $\protect\chi $}

The full response function $\chi $ is computed by including the Hartree
vertex corrections which, from Eq.(\ref{chi000}), gives 
\begin{align}
\chi _{n_{1},n_{2},n_{3},n_{4}}^{\alpha ,\beta ,\gamma ,\delta }\left( {\bf k%
},\omega \right) & =\widetilde{\chi }_{n_{1},n_{2},n_{3},n_{4}}^{\alpha
,\beta ,\beta ,\alpha }\left( {\bf k},\omega \right) \delta _{\alpha ,\delta
}\delta _{\beta ,\gamma }  \label{chi10} \\
& +\delta _{\alpha ,\beta }\sum_{n_{5}\cdots n_{8}}\sum_{\eta }\widetilde{%
\chi }_{n_{1},n_{2},n_{5},n_{6}}^{\alpha ,\alpha ,\alpha ,\alpha }\left( 
{\bf k},\omega \right) H_{n_{5},n_{6},n_{7},n_{8}}\left( {\bf k}\right) \chi
_{n_{7},n_{8},n_{3},n_{4}}^{\eta ,\eta ,\gamma ,\delta }\left( {\bf k}%
,\omega \right) ,  \nonumber
\end{align}
where 
\begin{equation}
\left[ H_{_{n_{1},n_{2},n_{3},n_{4}}}\left( {\bf k}\right) \right] _{{\bf G},%
{\bf G}}\equiv H\left( n_{1},n_{2},n_{3},n_{4};{\bf k+G}\right) \delta _{%
{\bf G},{\bf G}^{\prime }}.
\end{equation}
To simplify this equation, we will again consider separately the case of
intra and inter-Landau-level excitations.

For intra-Landau-level excitations, we write

\begin{equation}
\chi _{n,n,m,m}^{\alpha ,\beta ,\gamma ,\delta }\left( {\bf k},\omega
\right) =\widetilde{\chi }_{n,n,m,m}^{\alpha ,\beta ,\beta ,\alpha }\delta
_{\alpha ,\delta }\delta _{\beta ,\gamma }+\delta _{\alpha ,\beta
}\sum_{n_{5}\cdots n_{8}}\sum_{\eta }\widetilde{\chi }_{n,n,n_{5},n_{6}}^{%
\alpha ,\alpha ,\alpha ,\alpha }H_{n_{5},n_{6},n_{7},n_{8}}\left( {\bf k}%
\right) \chi _{n_{7},n_{8},m,m}^{\eta ,\eta ,\gamma ,\delta }\left( {\bf k}%
,\omega \right) .  \label{chi2}
\end{equation}
Using our approximation of no coupling between excitations of different $%
n\omega _{c}$, one may show that 
\begin{equation}
\left[ i\Omega _{n}I-\left( F_{p,p}^{\sigma ,\sigma }\left( {\bf k}\right)
+B_{p,p}^{\sigma ,\sigma }\left( {\bf k}\right) \left[ H_{p}^{(0)}\left( 
{\bf k}\right) -X_{p}^{(0)}\left( {\bf k}\right) \right] \right) \right]
\chi _{p,p,p,p}^{\sigma ,\sigma ,\sigma ,\sigma }\left( {\bf k},\omega
\right) =B_{p,p}^{\sigma ,\sigma }\left( {\bf k}\right) .  \label{phonon}
\end{equation}
where $\sigma $ is the spin index of the partially filled level and 
\begin{equation}
\left[ H_{n}^{(0)}\left( {\bf k}\right) \right] _{{\bf G},{\bf G}^{\prime
}}\equiv H\left( n,n,n,n;{\bf k+G}\right) \delta _{{\bf G},{\bf G}^{\prime
}}.
\end{equation}
The poles of $\chi _{p,p,p,p}^{\sigma ,\sigma ,\sigma ,\sigma }$ contain the
phonon mode. For spin waves, there is no Hartree vertex corrections and the
relevant response function obeys 
\begin{equation}
\left[ i\Omega _{n}I-\left( F_{p,p}^{+,-}\left( {\bf k}\right)
-B_{p,p}^{+,-}\left( {\bf k}\right) X_{n}^{(0)}\left( {\bf k}\right) \right) %
\right] \chi _{p,p,p,p}^{+,-,-,+}\left( {\bf k},\omega \right)
=B_{p,p}^{+,-}\left( {\bf k}\right) .  \label{spinwave}
\end{equation}

The most complex situation is that of inter-Landau-level density modes
(cyclotron modes). In this case, we need to consider the coupling between
Landau levels as well as between spins. From Eq. (\ref{chi10}) and Eq. (\ref
{spinflip}), we get 
\begin{align}
& \left[ i\Omega _{n}\overline{I}-F^{\alpha ,\alpha }\left( {\bf k}\right)
+B^{\alpha ,\alpha }\left( {\bf k}\right) \overline{V_{p}}\left( {\bf k}%
\right) \right] \overline{\chi }^{\alpha ,\alpha ,\beta ,\beta }\left( {\bf k%
},\omega \right)  \label{chi44} \\
& =\overline{B}^{\alpha ,\alpha }\left( {\bf k}\right) \delta _{\alpha
,\beta }+\overline{B}^{\alpha ,\alpha }\left( {\bf k}\right) \overline{H}%
_{n}\left( {\bf k}\right) \left[ \sum_{\eta }\overline{\chi }^{\eta ,\eta
,\beta ,\beta }\left( {\bf k},\omega \right) \right] ,  \nonumber
\end{align}
where 
\begin{equation}
\overline{\chi }^{\alpha ,\alpha ,\beta ,\beta }\equiv \left[ 
\begin{array}{cc}
\chi _{p,p+1,p+1,p}^{\alpha ,\alpha ,\beta ,\beta } & \chi
_{p,p+1,p,p-1}^{\alpha ,\alpha ,\beta ,\beta } \\ 
\chi _{p-1,p,p+1,p}^{\alpha ,\alpha ,\beta ,\beta } & \chi
_{p-1,p,p,p-1}^{\alpha ,\alpha ,\beta ,\beta }
\end{array}
\right]
\end{equation}
and 
\begin{equation}
\overline{H}_{n}\equiv \left[ 
\begin{array}{cc}
H_{n}^{(1)} & H_{n}^{(4)} \\ 
H_{n}^{(2)} & H_{n}^{(3)}
\end{array}
\right] ,
\end{equation}
with 
\begin{eqnarray}
\left[ H_{n}^{(1)}\left( {\bf k}\right) \right] _{{\bf G},{\bf G}^{\prime }}
&\equiv &H\left( n+1,n,n,n+1;{\bf k+G}\right) \delta _{{\bf G},{\bf G}%
^{\prime }};  \nonumber \\
\left[ H_{n}^{(2)}\left( {\bf k}\right) \right] _{{\bf G},{\bf G}^{\prime }}
&\equiv &H\left( n,n-1,n,n+1;{\bf k+G}\right) \delta _{{\bf G},{\bf G}%
^{\prime }}; \\
\left[ H_{n}^{(3)}\left( {\bf k}\right) \right] _{{\bf G},{\bf G}^{\prime }}
&\equiv &H\left( n,n-1,n-1,n;{\bf k+G}\right) \delta _{{\bf G},{\bf G}%
^{\prime }};  \nonumber \\
\left[ H_{n}^{(4)}\left( {\bf k}\right) \right] _{{\bf G},{\bf G}^{\prime }}
&\equiv &H\left( n+1,n,n-1,n;{\bf k+G}\right) \delta _{{\bf G},{\bf G}%
^{\prime }}.  \nonumber
\end{eqnarray}
Eq. (\ref{chi44}) can be written in a more transparent form by defining the
block matrices 
\begin{eqnarray}
\overline{\overline{\chi }} &\equiv &\left[ 
\begin{array}{cc}
\overline{\chi }^{+,+,+,+} & \overline{\chi }^{+,+,-,-} \\ 
\overline{\chi }^{-,-,+,+} & \overline{\chi }^{-,-,-,-}
\end{array}
\right]  \label{superchi} \\
&=&\left[ 
\begin{array}{cccc}
\chi _{p,p+1,p+1,p}^{+,+,+,+} & \chi _{p,p+1,p,p-1}^{+,+,+,+} & \chi
_{p,p+1,p+1,p}^{+,+,-,-} & \chi _{p,p+1,p,p-1}^{+,+,-,-} \\ 
\chi _{p-1,p,p+1,p}^{+,+,+,+} & \chi _{p-1,p,p,p-1}^{+,+,+,+} & \chi
_{p-1,p,p+1,p}^{+,+,-,-} & \chi _{p-1,p,p,p-1}^{+,+,-,-} \\ 
\chi _{p,p+1,p+1,p}^{-,-,+,+} & \chi _{p,p+1,p,p-1}^{-,-,+,+} & \chi
_{p,p+1,p+1,p}^{-,-,-,-} & \chi _{p,p+1,p,p-1}^{-,-,-,-} \\ 
\chi _{p-1,p,p+1,p}^{-,-,+,+} & \chi _{p-1,p,p,p-1}^{-,-,+,+} & \chi
_{p-1,p,p+1,p}^{-,-,-,-} & \chi _{p-1,p,p,p-1}^{-,-,-,-}
\end{array}
\right] ,  \nonumber
\end{eqnarray}
\begin{equation}
\overline{\overline{B}}\equiv \left[ 
\begin{array}{cc}
\overline{B}^{+,+} & \overline{0} \\ 
\overline{0} & \overline{B}^{-,-}
\end{array}
\right] =\left[ 
\begin{array}{cccc}
B_{p,p+1}^{+,+} & 0 & 0 & 0 \\ 
0 & B_{p-1,p}^{+,+} & 0 & 0 \\ 
0 & 0 & B_{p,p+1}^{-,-} & 0 \\ 
0 & 0 & 0 & B_{p-1,p}^{-,-}
\end{array}
\right] ,
\end{equation}
\begin{eqnarray}
\overline{\overline{V}}_{p} &\equiv &\left[ 
\begin{array}{cc}
\overline{H}_{p}-\overline{X}_{p} & \overline{H}_{p} \\ 
\overline{H}_{p} & \overline{H}_{p}-\overline{X}_{p}
\end{array}
\right] \\
&=&\left[ 
\begin{array}{cccc}
H_{p}^{(1)}-X_{p}^{(1)} & H_{p}^{(4)}-X_{p}^{(4)} & H_{p}^{(1)} & H_{p}^{(4)}
\\ 
H_{p}^{(2)}-X_{p}^{(2)} & H_{p}^{(3)}-X_{p}^{(3)} & H_{p}^{(2)} & H_{p}^{(3)}
\\ 
H_{p}^{(1)} & H_{p}^{(4)} & H_{p}^{(1)}-X_{p}^{(1)} & H_{p}^{(4)}-X_{p}^{(4)}
\\ 
H_{p}^{(2)} & H_{p}^{(3)} & H_{p}^{(2)}-X_{p}^{(2)} & H_{p}^{(3)}-X_{p}^{(3)}
\end{array}
\right] ,
\end{eqnarray}
\begin{equation}
\overline{\overline{I}}\equiv \left[ 
\begin{array}{cc}
\overline{I} & \overline{0} \\ 
\overline{0} & \overline{I}
\end{array}
\right] =\left[ 
\begin{array}{cccc}
I & 0 & 0 & 0 \\ 
0 & I & 0 & 0 \\ 
0 & 0 & I & 0 \\ 
0 & 0 & 0 & I
\end{array}
\right],
\end{equation}
\begin{equation}
\overline{\overline{F}}\equiv \left[ 
\begin{array}{cc}
\overline{F}^{++} & \overline{0} \\ 
\overline{0} & \overline{F}^{--}
\end{array}
\right] =\left[ 
\begin{array}{cccc}
F_{p,p+1}^{+,+}\left( {\bf k}\right) & 0 & 0 & 0 \\ 
0 & F_{p-1,p}^{+,+}\left( {\bf k}\right) & 0 & 0 \\ 
0 & 0 & F_{p,p+1}^{-,-}\left( {\bf k}\right) & 0 \\ 
0 & 0 & 0 & F_{p-1,p}^{-,-}\left( {\bf k}\right)
\end{array}
\right] .
\end{equation}
The equation of motion then takes the form 
\begin{equation}
\left[ i\Omega _{n}\overline{\overline{I}}-\overline{\overline{F}}\left( 
{\bf k}\right) -\overline{\overline{B}}\left( {\bf k}\right) \overline{%
\overline{V}}_{p}\left( {\bf k}\right) \right] \overline{\overline{\chi }}%
\left( {\bf k,}\omega \right) =\overline{\overline{B}}\left( {\bf k}\right)
\label{cyclotron}
\end{equation}
We solve this matrix equation numerically to obtain response functions whose
poles give the magnetoplasmon and inter-Landau-level spin-flip excitations.

\section{Dispersion relations in the liquid phase}

\label{app:liquid}

The equations of the previous section can be drastically simplified in the
homogeneous or liquid phase since then $\left\langle \rho _{n}^{\alpha
}\left( {\bf G}\right) \right\rangle =\widetilde{\nu }\delta _{{\bf G},0}$
or $\left\langle \rho _{n}^{\alpha }\left( {\bf G}\right) \right\rangle =0$.
For example, for $0\leq \nu \leq 2$, the dispersion relations for various
modes can be computed analytically, and one may show they reproduce the
results of Kallin and Halperin\cite{kallin}. Larger filling factors are more
complicated as they involve several different particle-hole excitations; the
coupling among these has not been treated correctly in previous studies\cite
{kallin}. As concrete examples of the present method, we compute the
collective modes for the liquid state at different filling factors and, in
particular, for $\nu =6.45$ which corresponds to the filling factor of the
stripe crystal considered in this paper.

\subsubsection{Liquid phase with $0\leq \protect\nu \leq 1$}

As an application of the above formalism, we consider here the simple case
of $0\leq \nu \leq 1.$ If the lowest Landau level is partially occupied with
up spins, then $\left\langle \rho _{0}^{+}\left( {\bf G}\right)
\right\rangle =\nu \delta _{{\bf G},0}$. All matrices are diagonal and so $%
{\bf k+G\rightarrow q}$ which is not restricted to the first Brillouin zone.

Since $B_{0,0}^{+,+}\left( {\bf q}\right) =0$ if follows that there can be
no phonon mode. Moreover, since $B_{0,0}^{+,-}\left( {\bf q}\right) =\nu $,
we have from Eq.(\ref{spinwave}) 
\begin{equation}
\omega _{SW}\left( {\bf q}\right) =g^{\ast }\mu _{B}B+\nu \left[
X_{0,0}\left( 0\right) -X_{0}^{(0)}\left( {\bf q}\right) \right] .
\end{equation}
Since $X_{0}^{(0)}\left( {\bf q}=0\right) =X_{0,0}\left( 0\right) ,$ it
follows that $\omega _{SW}\left( {\bf 0}\right) =g^{\ast }\mu _{B}B$ as
required by Larmor's theorem.

For the density mode $B_{0,1}^{+,+}\left( {\bf q}\right) =\nu $, and the
dispersion is 
\begin{equation}
\omega _{nn}\left( {\bf q}\right) =\omega _{c}+\nu \left[ X_{0,0}\left(
0\right) -X_{1,0}(0)+H_{1}^{(0)}\left( {\bf q}\right) -X_{1}^{(0)}\left( 
{\bf q}\right) \right] .
\end{equation}
In this equation $\nu X_{0,0}\left( 0\right) $ is the self-energy lost by
the electron leaving level $n=0$ while $\nu X_{1,0}\left( 0\right) $ is the
self-energy gained in the new level $n=1$. Because $H_{1}^{(0)}\left(
0\right) =0$ and $X_{0,0}\left( 0\right) -X_{1,0}(0)-X_{1}^{(0)}\left(
0\right) =0,$ it follows that $\omega _{{\rm cyc}.}\left( 0\right) =\omega
_{c}$ as required by Kohn's theorem. We remark that these two results are
identical to those of Ref. \cite{kallin}

For the inter-Landau-level spin flip excitation, $B_{0,1}^{+,-}\left( {\bf q}%
\right) =\nu $ and the dispersion is 
\begin{equation}
\omega _{SF}\left( {\bf q}\right) =\omega _{c}+g^{\ast }\mu _{B}B+\nu \left[
X_{0,0}\left( 0\right) -X_{1}^{(0)}\left( {\bf q}\right) \right] .
\end{equation}
Notice that $\left[ X_{0,0}\left( 0\right) -X_{1}^{(0)}(0)\right] >0$ so
that the self-energy and vertex correction introduce a positive shift in the
dispersion relation contrary to the result in Ref. \cite{kallin}. This was
first noticed in Ref. \cite{brey}. Similar problems with the inclusion of
the self-energy terms appear in the higher-energy modes as well in Ref. \cite
{kallin}. Apart from this discrepancy, our results reproduces correctly the
dispersion relation of the higher-energy modes of the liquid phase.

\subsubsection{Liquid phase with $6\leq \protect\nu \leq 7$}

This is the case we consider in the stripe phase. It is thus interesting to
compare the dispersion relations obtained there with the corresponding ones
in the liquid phase. We assume that the partially filled level is $(p=3,+)$,
so that $\left\langle \rho _{3}^{+}\left( {\bf G}\right) \right\rangle =%
\widetilde{\nu }\delta _{{\bf G},0}$ and $\left\langle \rho _{m}^{\alpha
}\left( {\bf G}\right) \right\rangle =\delta _{{\bf G},0}$ for $m<3$. There
is again, of course, no phonon mode.

For spin wave, $B_{3,3}^{+,-}\left( {\bf q}\right) =\widetilde{\nu }$ and
the dispersion, from Eq. (\ref{spinwave}) is simply 
\begin{equation}
\omega _{SW}\left( {\bf q}\right) =g^{\ast }\mu _{B}B+\widetilde{\nu }\left[
X_{3,3}\left( 0\right) -X_{3}^{(0)}\left( {\bf q}\right) \right] ,
\end{equation}
with $\omega _{SW}\left( 0\right) =g^{\ast }\mu _{B}B$ as required.

For spin-flip excitations with $\delta S_{z}=+1$, we must look at $\chi
_{2,3,3,2}^{-,+,+,-}$ which, from Eq. (\ref{spinflip}) is coupled to $\chi
_{3,4,3,2}^{-,+,+,-}$. Solving this system of equation, we rapidly obtain
that $\chi _{3,4,4,3}^{-,+,+,-}=0$ so that the dispersion relation is
obtained from $\chi _{2,3,3,2}^{-,+,+,-}$ only. This makes sense, since the
transition $\left( 2,-\right) \rightarrow \left( 3,+\right) $ is not coupled
to any other in the situation we consider. We find then 
\begin{equation}
\omega _{SF+}\left( {\bf k}\right) =\omega _{c}-g^{\ast }\mu _{B}B+\Sigma
_{2,3}^{-,+}-\left( 1-\widetilde{\nu }\right) X_{3}^{(3)}\left( {\bf k}%
\right) ,
\end{equation}
where 
\begin{equation}
\Sigma _{2,3}^{-,+}\equiv X_{2,0}\left( 0\right) +X_{2,1}\left( 0\right)
+X_{2,2}\left( 0\right) -X_{3,0}\left( 0\right) -X_{3,1}\left( 0\right)
-X_{3,2}\left( 0\right) -\widetilde{\nu }X_{3,3}\left( 0\right) .
\end{equation}
The exchange and vertex corrections introduce a downward shift in $\omega
_{SF+}(0)$ from $\omega _{c}-g^{\ast }\mu _{B}B$ (see Fig. 6).

For spin-flip excitations with $\delta S_{z}=-1,$ the transition $\left(
2,+\right) \rightarrow \left( 3,-\right) $ is coupled to $\left( 3,+\right)
\rightarrow \left( 4,-\right) $ and we must solve Eq.(\ref{spinflip}) that
couples $\chi _{3,4,4,3}^{+,-,-,+}$ to $\chi _{2,3,4,3}^{+,-,-,+}.$ There is
correspondingly two such spin-flip modes, with dispersion given by 
\begin{align}
\omega _{SF-}\left( {\bf k}\right) & =\omega _{c}+g^{\ast }\mu _{B}B{\bf +}%
\frac{1}{2}\left( \Lambda _{2,3}^{+,-}\left( {\bf k}\right) +\Lambda
_{3,4}^{+,-}\left( {\bf k}\right) \right) \\
& \pm \sqrt{\left( \Lambda _{2,3}^{+,-}\left( {\bf k}\right) +\Lambda
_{3,4}^{+,-}\left( {\bf k}\right) \right) ^{2}-4\left[ \Lambda
_{2,3}^{+,-}\left( {\bf k}\right) \Lambda _{3,4}^{+,-}\left( {\bf k}\right) -%
\widetilde{\nu }X_{3}^{(4)}\left( {\bf k}\right) X_{3}^{(2)}\left( {\bf k}%
\right) \right] },  \nonumber
\end{align}
where 
\begin{eqnarray}
\Lambda _{2,3}^{+,-}\left( {\bf k}\right) &\equiv &\Sigma _{2,3}^{+,-}-%
\widetilde{\nu }X_{3}^{(1)}\left( {\bf k}\right) , \\
\Lambda _{3,4}^{+,-}\left( {\bf k}\right) &\equiv &\Sigma
_{3,4}^{+,-}-X_{3}^{(3)}\left( {\bf k}\right) ,
\end{eqnarray}
with 
\begin{eqnarray}
\Sigma _{2,3}^{+,-} &\equiv &X_{2,0}\left( 0\right) +X_{2,1}\left( 0\right)
+X_{2,2}\left( 0\right) +\widetilde{\nu }X_{2,3}\left( 0\right)
-X_{3,0}\left( 0\right) -X_{3,1}\left( 0\right) -X_{3,2}\left( 0\right) , \\
\Sigma _{3,4}^{+,-} &\equiv &X_{3,0}\left( 0\right) +X_{3,1}\left( 0\right)
+X_{3,2}\left( 0\right) +\widetilde{\nu }X_{3,3}\left( 0\right)
-X_{4,0}\left( 0\right) -X_{4,1}\left( 0\right) -X_{4,2}\left( 0\right) .
\end{eqnarray}
In this case, the shift is positive in both modes (see Fig. 6).

For the density modes, three excitations are coupled: $\left( 2,+\right)
\rightarrow \left( 3,+\right) ,\left( 2,-\right) \rightarrow \left(
3,-\right) $ and $\left( 3,+\right) \rightarrow \left( 4,+\right) $. Since $%
B_{3,4}^{-,-}\left( {\bf k}\right) =0$, $\chi _{3,4,4,3}^{-,-,-,-}\left( 
{\bf k},\omega \right) =0$ and the $4\times 4$ block matrix in Eq.(\ref
{superchi}) reduces to a $3\times 3$ block matrix. The three collective
modes are found from the determinant of $\left[ (\omega +i\delta )\overline{%
\overline{I}}-\overline{\overline{F}}\left( {\bf k}\right) -\overline{%
\overline{B}}\left( {\bf k}\right) \overline{\overline{V}}_{p}\left( {\bf k}%
\right) \right] $ in Eq.(\ref{cyclotron}) i.e. from 
\begin{equation}
\left| \left[ 
\begin{array}{ccc}
\left( \omega -\omega _{c}\right) -\Lambda _{3,4}^{+,+}\left( {\bf k}\right)
& -\widetilde{\nu }\left( H_{3}^{(2)}\left( {\bf k}\right)
-X_{3}^{(2)}\left( {\bf k}\right) \right) & -\widetilde{\nu }%
H_{3}^{(2)}\left( {\bf k}\right) \\ 
-\left( 1-\widetilde{\nu }\right) \left( H_{3}^{(2)}\left( {\bf k}\right)
-X_{3}^{(2)}\left( {\bf k}\right) \right) & \left( \omega -\omega
_{c}\right) -\Lambda _{2,3}^{+,+}\left( {\bf k}\right) & -\left( 1-%
\widetilde{\nu }\right) H_{3}^{(3)}\left( {\bf k}\right) \\ 
-H_{3}^{(2)}\left( {\bf k}\right) & -H_{3}^{(3)}\left( {\bf k}\right) & 
\left( \omega -\omega _{c}\right) -\Lambda _{2,3}^{-,-}\left( {\bf k}\right)
\end{array}
\right] \right| =0,
\end{equation}
where 
\begin{eqnarray}
\Lambda _{3,4}^{+,+}\left( {\bf k}\right) &\equiv &\Sigma _{3,4}^{+,+}+%
\widetilde{\nu }\left( H_{3}^{(1)}\left( {\bf k}\right) -X_{3}^{(1)}\left( 
{\bf k}\right) \right) , \\
\Lambda _{2,3}^{+,+}\left( {\bf k}\right) &\equiv &\Sigma
_{2,3}^{+,+}+\left( 1-\widetilde{\nu }\right) \left( H_{3}^{(3)}\left( {\bf k%
}\right) -X_{3}^{(3)}\left( {\bf k}\right) \right) , \\
\Lambda _{2,3}^{-,-}\left( {\bf k}\right) &\equiv &\Sigma
_{2,3}^{-,-}+\left( H_{3}^{(3)}\left( {\bf k}\right) -X_{3}^{(3)}\left( {\bf %
k}\right) \right) ,
\end{eqnarray}
and 
\begin{align}
\Sigma _{3,4}^{+,+}& \equiv \Sigma _{3,4}^{+,-}-\widetilde{\nu }%
X_{4,3}\left( 0\right) , \\
\Sigma _{2,3}^{+,+}& \equiv \Sigma _{2,3}^{+,-}-\widetilde{\nu }%
X_{3,3}\left( 0\right) , \\
\Sigma _{2,3}^{-,-}& \equiv X_{2,0}\left( 0\right) +X_{2,1}\left( 0\right)
+X_{2,2}\left( 0\right) -X_{3,0}\left( 0\right) -X_{3,1}\left( 0\right)
-X_{3,2}\left( 0\right) .
\end{align}

These collectives modes are represented in Fig. 5.

\section{Harmonic theory}

\label{harmonic}

As discussed in the text, to an excellent approximation the phonon mode
frequencies computed in the TDHFA for the stripe phase disperse linearly
from $k_{\parallel }=0$, with a slope that vanishes at $k_{\perp }=0$. In
this Appendix we demonstrate that this behavior is consistent with a
harmonic theory for a two-dimensional charged smectic system in a magnetic
field. Defining the direction parallel to the stripe as the $\hat{x}$
direction, the simplest long-wavelength harmonic potential one can write
down might be 
\begin{equation}
V={\frac{1}{2}}\int d^{2}r\left[ \kappa _{x}\left( {\frac{{\partial u^{x}}}{{%
\partial x}}}\right) ^{2}+\kappa _{y}\left( {\frac{{\partial u^{y}}}{{%
\partial y}}}\right) ^{2}\right] .  \label{simple}
\end{equation}
In the above equation, ${\bf u}$ is a displacement field for the stripes,
the first term represents an elastic contribution for longitudinal
compression of the stripes, and the second arises from interstripe
repulsion. Collective modes are most easily computed in terms of the Fourier
transformed displacements, 
\begin{equation}
{\bf u}({\bf q})={\frac{1}{\sqrt{A}}}\int d^{2}r\,e^{i{\bf q}\cdot {\bf r}}%
{\bf u}\left( {\bf r}\right) ,
\end{equation}
where $A$ is the area of the system. To compute the collective modes in a
single Landau level, one may impose the commutation relations\cite{vignale} $%
\left[ u^{x}\left( {\bf q}_{1}\right) ,u^{y}\left( {\bf q}_{2}\right) \right]
=i\ell ^{2}\delta _{{\bf q}_{1},-{\bf q}_{2}}$. The equation of motion $i%
\frac{du_{{\bf q}}^{\mu }}{dt}=\left[ u_{{\bf q}}^{\mu },V\right] $, where $%
\mu =x,y$, after Fourier transform with respect to time may be written in
the form 
\begin{equation}
-i\ell ^{2}\left( 
\begin{array}{cc}
D_{yx}\left( {\bf q}\right) & D_{yy}\left( {\bf q}\right) \\ 
-D_{xx}\left( {\bf q}\right) & -D_{xy}\left( {\bf q}\right)
\end{array}
\right) \left( 
\begin{array}{c}
u^{x}\left( {\bf q}\right) \\ 
u^{y}\left( {\bf q}\right)
\end{array}
\right) =\omega \left( 
\begin{array}{c}
u^{x}\left( {\bf q}\right) \\ 
u^{y}\left( {\bf q}\right)
\end{array}
\right) .  \label{dynmat}
\end{equation}
For a system with inversion symmetry, $D_{xy}\left( {\bf q}\right)
=D_{yx}\left( {\bf q}\right) $, and the eigenvalues of Eq. (\ref{dynmat})
(i.e., the collective mode frequencies) take the general form 
\begin{equation}
\omega \left( {\bf q}\right) =\pm \ell ^{2}\sqrt{D_{xx}\left( {\bf q}\right)
D_{yy}\left( {\bf q}\right) -|D_{xy}\left( {\bf q}\right) |^{2}}.
\label{eval}
\end{equation}
For Eq. (\ref{simple}), $D_{\mu ,\mu }=\kappa _{\mu }q_{\mu }^{2},~D_{xy}=0$%
, so $\omega \left( {\bf q}\right) \propto q_{y}q_{x}$. This has the correct
behavior of linear dispersion with respect to $q_{x}$, with a slope that
vanishes as $q_{y}\rightarrow 0$. However, the model has the incorrect
behavior of containing zero modes along {\it both} the $q_{x}$ and $q_{y}$
axes\cite{com3}.

The key missing ingredient in Eq.(\ref{simple}) is that the restoring force
for motion perpendicular to the stripes comes only from interstripe
repulsion. We expect, however, that individual stripes resist bending, as in
a smectic system. Thus, one should add a curvature term to the energy, which
in this case we take to be 
\begin{equation}
V_{{\rm bend}}={\frac{1}{2}}\kappa _{b}\int d^{2}r\left( {\frac{{d^{2}u^{y}}%
}{{dx^{2}}}}\right) ^{2}.
\end{equation}
Adding this to $V$, the resulting $D_{yy}\left( {\bf q}\right) $ is modified
to $\kappa _{y}q_{y}^{2}+\kappa _{b}q_{x}^{4}$. The collective mode
frequencies then take the form $\omega \left( {\bf q}\right) \propto q_{x}%
\sqrt{q_{y}^{2}+\left( {\frac{{\kappa _{b}}}{{\kappa _{y}}}}\right) q_{x}^{4}%
}$. This has the correct behavior that collective modes are gapped except
for $q_{x}=0$. A further refinement necessary to correctly describe the
long-wavelength physics of this system is the addition of the Coulomb
interaction. This may be simply modeled by adding a term of the form 
\begin{equation}
V_{{\rm Coul}}={\frac{1}{2}}\kappa _{c}\sum_{{\bf q}}{\frac{\left| {{\bf q}%
\cdot {\bf u}\left( {\bf q}\right) }\right| {^{2}}}{q}}
\end{equation}
with $\kappa _{c}=2\pi e^{2}/\kappa a_{c}^{2}$, where in this last
expression $\kappa $ is the dielectric constant of the host material, and $%
a_{c}$ is the area per electron in the groundstate. Finally, symmetry also
allows the addition of a term of the form 
\begin{equation}
V_{xy}={\frac{1}{2}}\kappa _{xy}\int d^{2}r\left( {\frac{{du^{x}}}{{dx}}}{%
\frac{{du^{y}}}{{dy}}}\right) .
\end{equation}
Taking our potential energy to be $V+V_{bend}+V_{Coul}+V_{xy}$, one may
easily compute $D_{\mu ,\nu }$ and use Eq.(\ref{eval}) to show $\omega
\left( {\bf q}\right) \propto q_{x}$ for $q_{y}>0$, and $\omega \left( {\bf q%
}\right) \propto q_{x}^{5/2}$ for $q_{y}=0$. The sublinear behavior of the
collective mode at $q_{y}=0$ is consistent with the results of the TDHFA.

\newpage

\begin{center}
{\bf Figures Captions\bigskip }
\end{center}

\begin{enumerate}
\item[Fig.1]  Electron density for a stripe state with $\nu =6.45$. The
separation between the stripes is $a=7.16\ell $ and the period of the
modulations along the stripes is $b=1.95\ell .$ The modulations on two
adjacent stripes are displaced by $b/2$. This pattern can be described by a
primitive unit cell with lattice vectors ${\bf R}_{1}=$ $\left( a,b/2\right) 
$ and ${\bf R}_{2}=\left( 0,b\right) .$

\item[Fig. 2]  Phonon dispersion relation of the stripe phase with $\nu
=6.45 $. Left inset: phonon dispersion along the stripes for different
values of the wave vector $k_{\bot }$. The local minimum has a frequency of $%
\omega \approx 0.02$ $\left( e^{2}/\kappa \ell \right) $ and becomes soft as
filling factor is decreased. Right inset: Brillouin zone for the primitive
unit cell described in Fig. 1.

\item[Fig. 3]  Phonon density of states. (The small oscillations are
numerical artifacts).

\item[Fig. 4]  Dispersion relation $\omega -\omega _{c}$ of the three
branches of the magnetoplasmon mode in the stripe phase along the direction
perpendicular to the stripes in the density pattern of Fig. 1. The
corresponding dispersions in the liquid phase (see Fig. 5) have been folded
in the first Brillouin zone and are represented by full lines with the heavy
lines indicating parts of these dispersions that lie in the first Brillouin
zone. The filled circles represent frequencies obtained from $\chi _{nn}$
while the empty squares represent frequencies obtained from $\chi _{\sigma
_{z}}.$

\item[Fig. 5]  Dispersion relations $\omega -\omega _{c}$ of the three
magnetoplasmon modes of the liquid phase with $\nu =6.45.$

\item[Fig. 6]  Dispersion relations of the spin wave and spin flip modes for 
$\nu =6.45$ in the liquid phase. The dispersion $\omega _{SW}-g^{\ast }\mu
_{B}B$ of the intra-Landau-level spin wave mode is represented by the dashed
line. The dispersions $\omega _{SF-}-\left( \omega _{c}+g^{\ast }\mu
_{B}B\right) $ of the two branches of the inter-Landau-level spin flip mode
with $\delta S_{z}=-1$ are represented by the full lines. The dispersion $%
\omega _{SF+}-\left( \omega _{c}-g^{\ast }\mu _{B}B\right) $ of the
inter-Landau-level spin flip mode with $\delta S_{z}=+1$ is represented by
the dot-dashed line.

\item[Fig. 7]  Dispersion relation $\omega _{SW}-g^{\ast }\mu _{B}B$ of the
spin wave mode in the stripe phase along the direction perpendicular to the
stripes in the density pattern of Fig. 1. The corresponding dispersion in
the liquid phase (see Fig. 6) has been folded in the first Brillouin zone
and is represented by full lines with the heavy line indicating parts of
these dispersions that lie in the first Brillouin zone.

\item[Fig. 8]  Dispersion relation $\omega _{SW}-g^{\ast }\mu _{B}B$ of the
spin wave mode in the stripe phase along the direction parallel to the
stripes in the density pattern of Fig. 1. The corresponding dispersion in
the liquid phase (see Fig. 6) has been folded in the first Brillouin zone
and is represented by full lines with the heavy line indicating parts of
these dispersions that lie in the first Brillouin zone.

\item[Fig. 9]  Dispersion relation $\omega -\omega _{c}$ of the three
branches of the magnetoplasmon mode in the stripe phase along the direction
parallel to the stripes in the density pattern of Fig. 1. The corresponding
dispersions in the liquid phase (see Fig. 5) have been folded in the first
Brillouin zone and are represented by full lines with the heavy lines
indicating parts of these dispersions that lie in the first Brillouin zone.
The filled circles represent frequencies obtained from $\chi _{nn}$ while
the empty squares represent frequencies obtained from $\chi _{\sigma _{z}}.$

\item[Fig. 10]  Dispersion relations $\omega _{SF-}-\left( \omega
_{c}+g^{\ast }\mu _{B}B\right) $ of the spin flip mode in the stripe phase
along the direction perpendicular to the stripes in the density pattern of
Fig. 1. The corresponding dispersion in the liquid phase (Fig. 6) has been
folded in the first Brillouin zone and is represented by full lines with the
heavy line indicating parts of these dispersions that lie in the first
Brillouin zone.

\item[Fig. 11]  Dispersion relations $\omega _{SF+}-\left( \omega
_{c}-g^{\ast }\mu _{B}B\right) $ of the spin flip mode in the stripe phase
along the direction parallel to the stripes in the density pattern of Fig.
1. The corresponding dispersion in the liquid phase (see Fig. 6) has been
folded in the first Brillouin zone and is represented by full lines with the
heavy line indicating parts of these dispersions that lie in the first
Brillouin zone.
\end{enumerate}


\begin{references}
\bibitem{modecom}  In principle, modes that become soft in this way can be
detected by inelastic light scattering \cite{pinczuk}. We note, however,
that our calculation of the phonon DOS indicates only a small weight for
this mode, which may make such detection difficult.
\bibitem{com2}  Our results are consistent with those of Ref. \cite{kallin},
apart from some exchange corrections terms that were missing in that work.
See Ref. \cite{brey}.
\bibitem{com3}  This model was employed in Ref. \cite{fertig}, where only
the behavior near $q_{x}=0$ was important. The extra line of zero modes near 
$q_{y}=0$ was not included in the analysis, nor should it have been since
this is only an artifact of Eq. \ref{simple}.
\end{references}
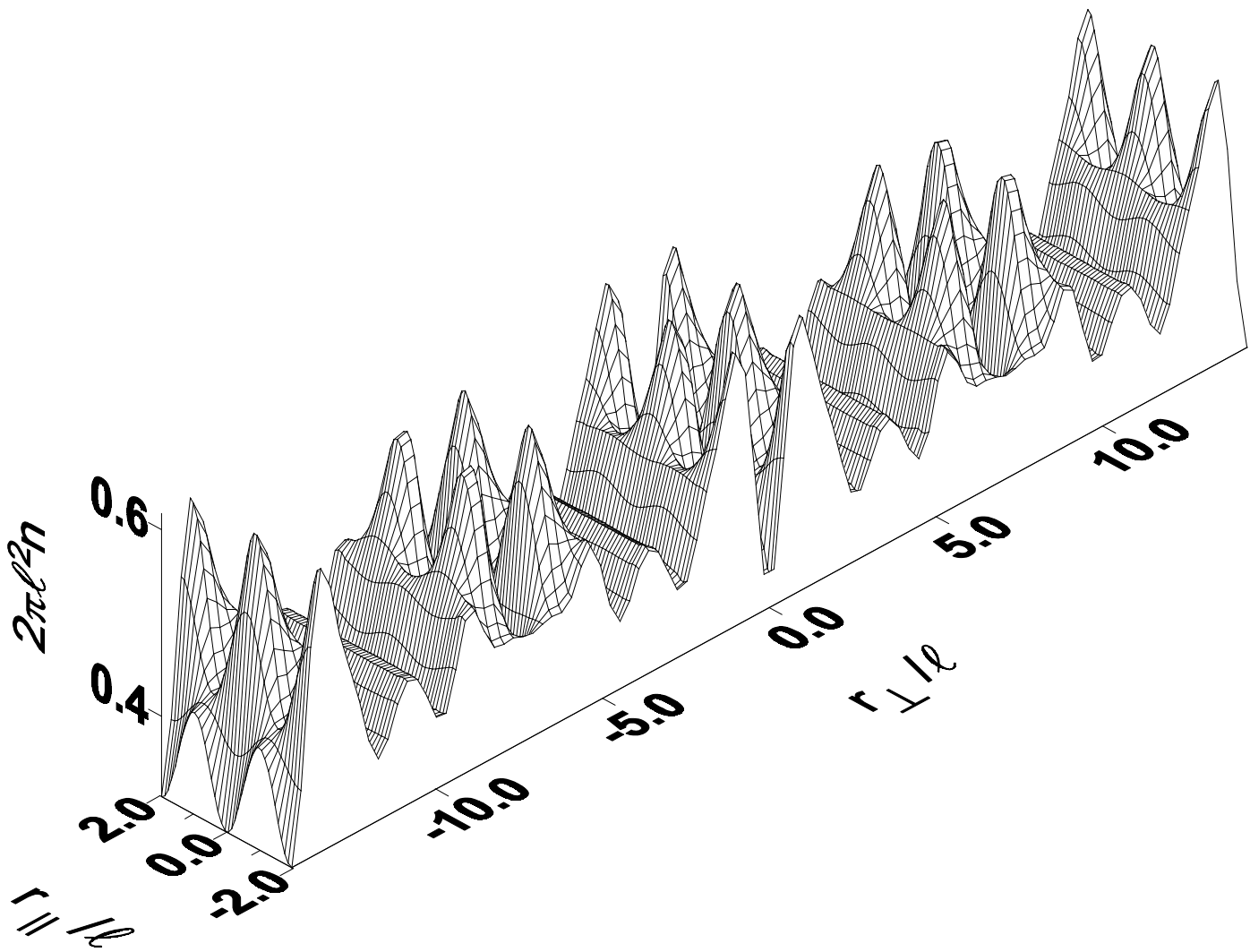
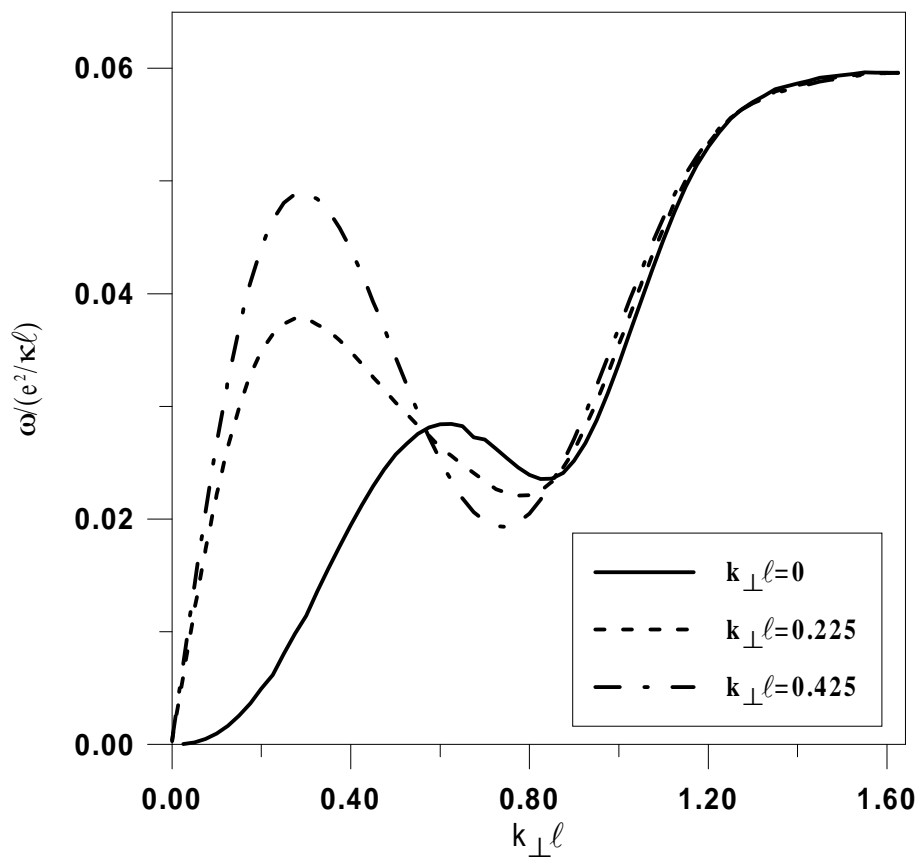
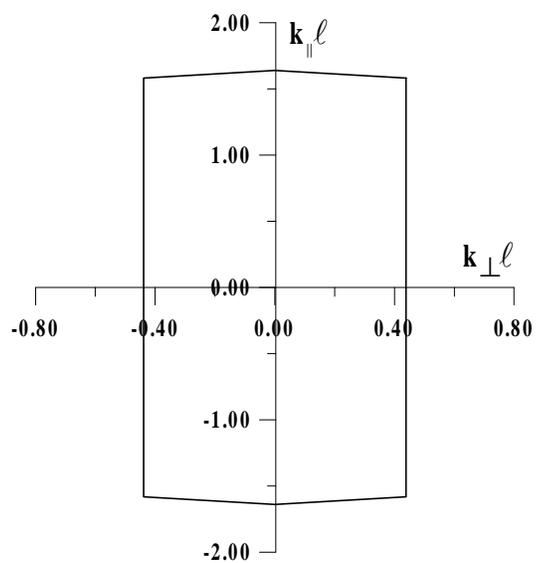
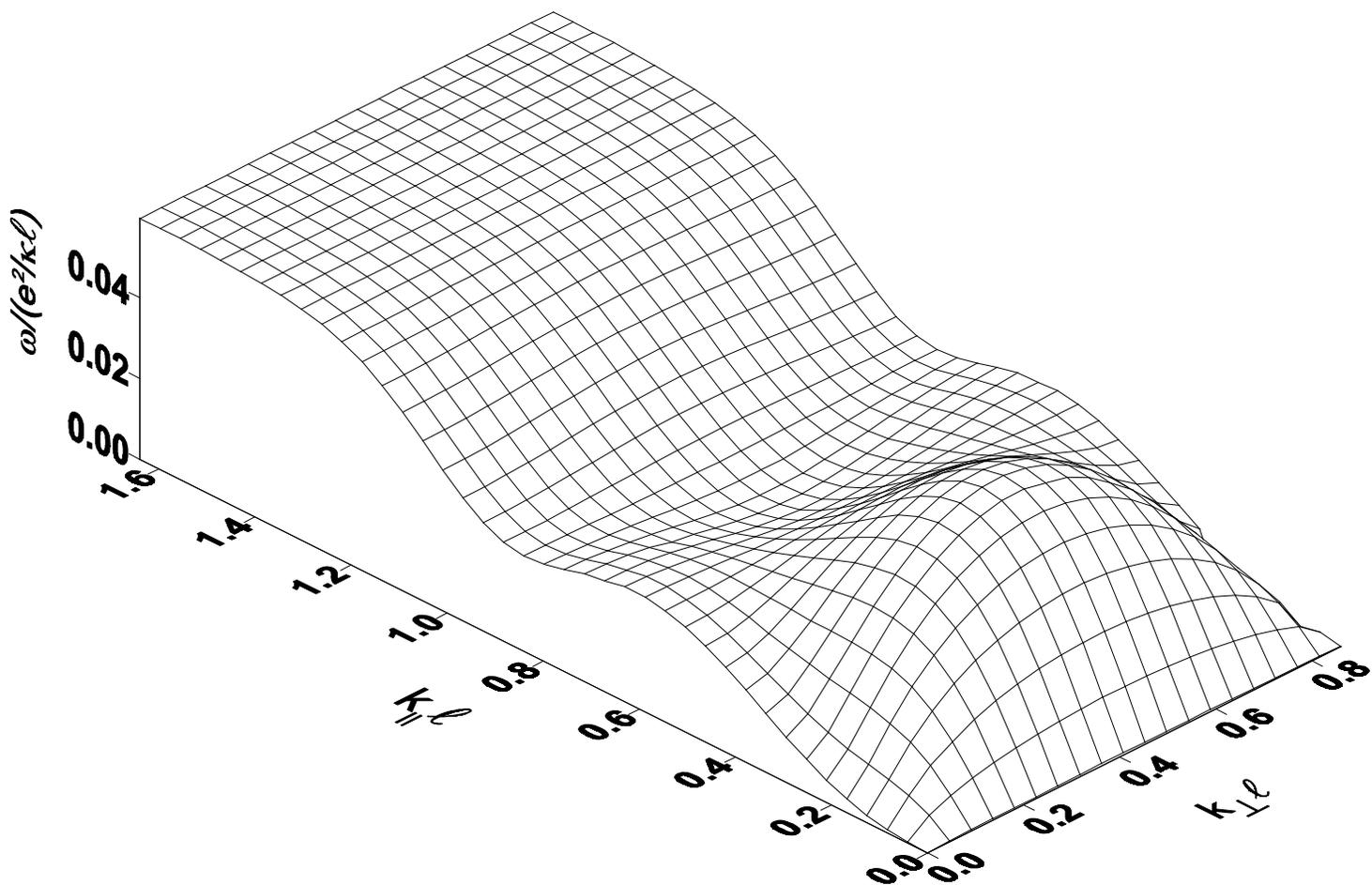
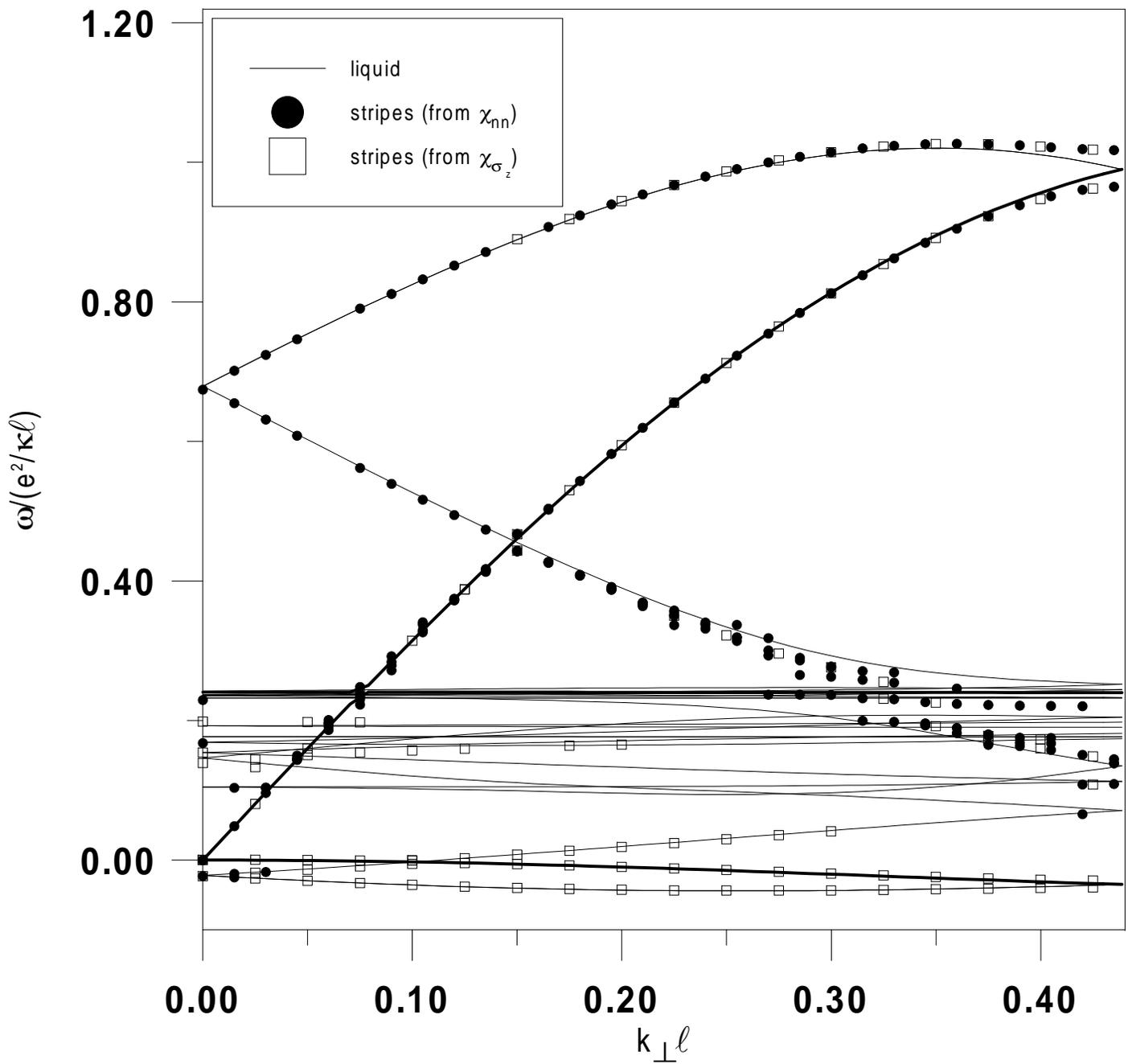
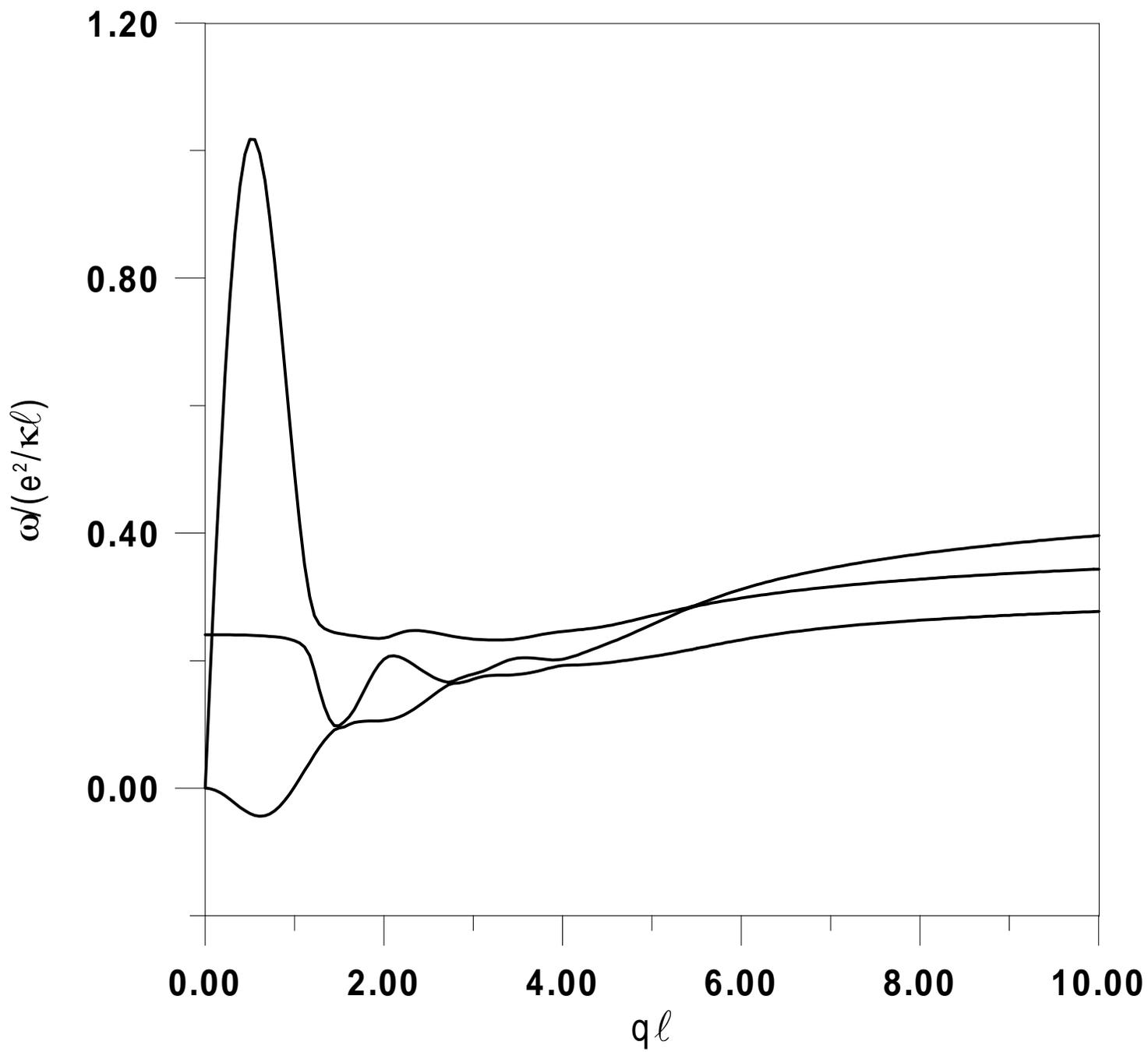
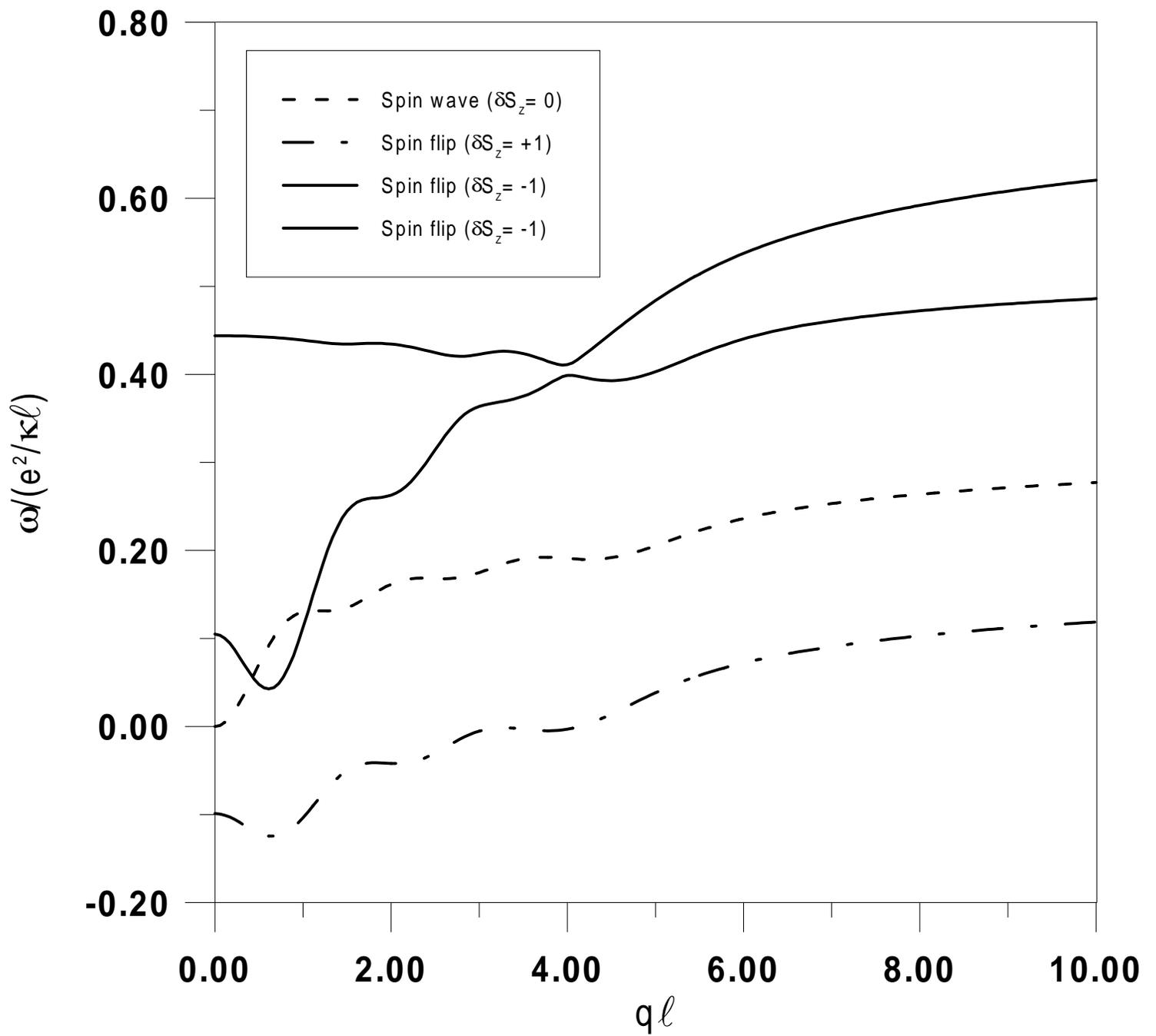
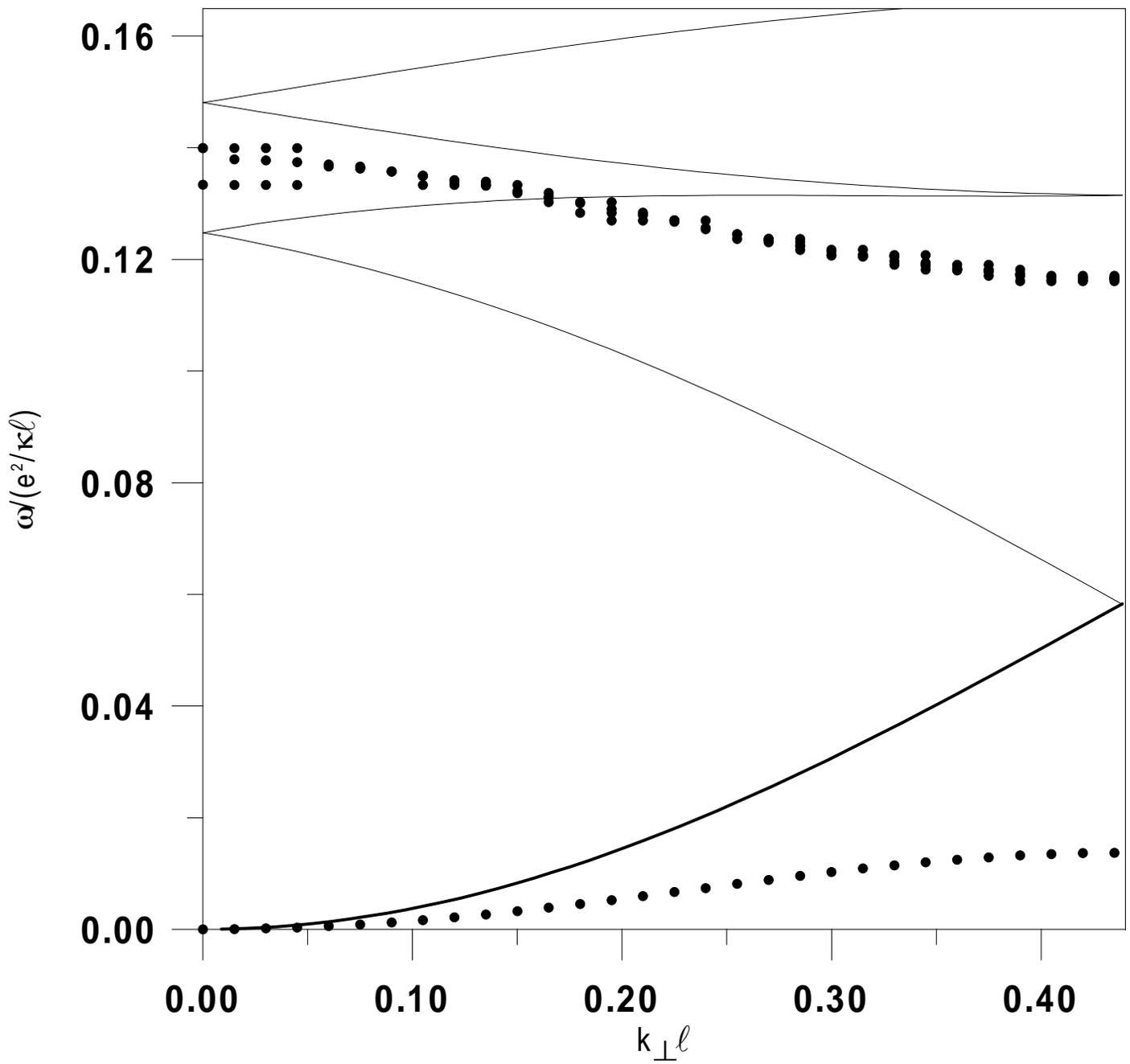
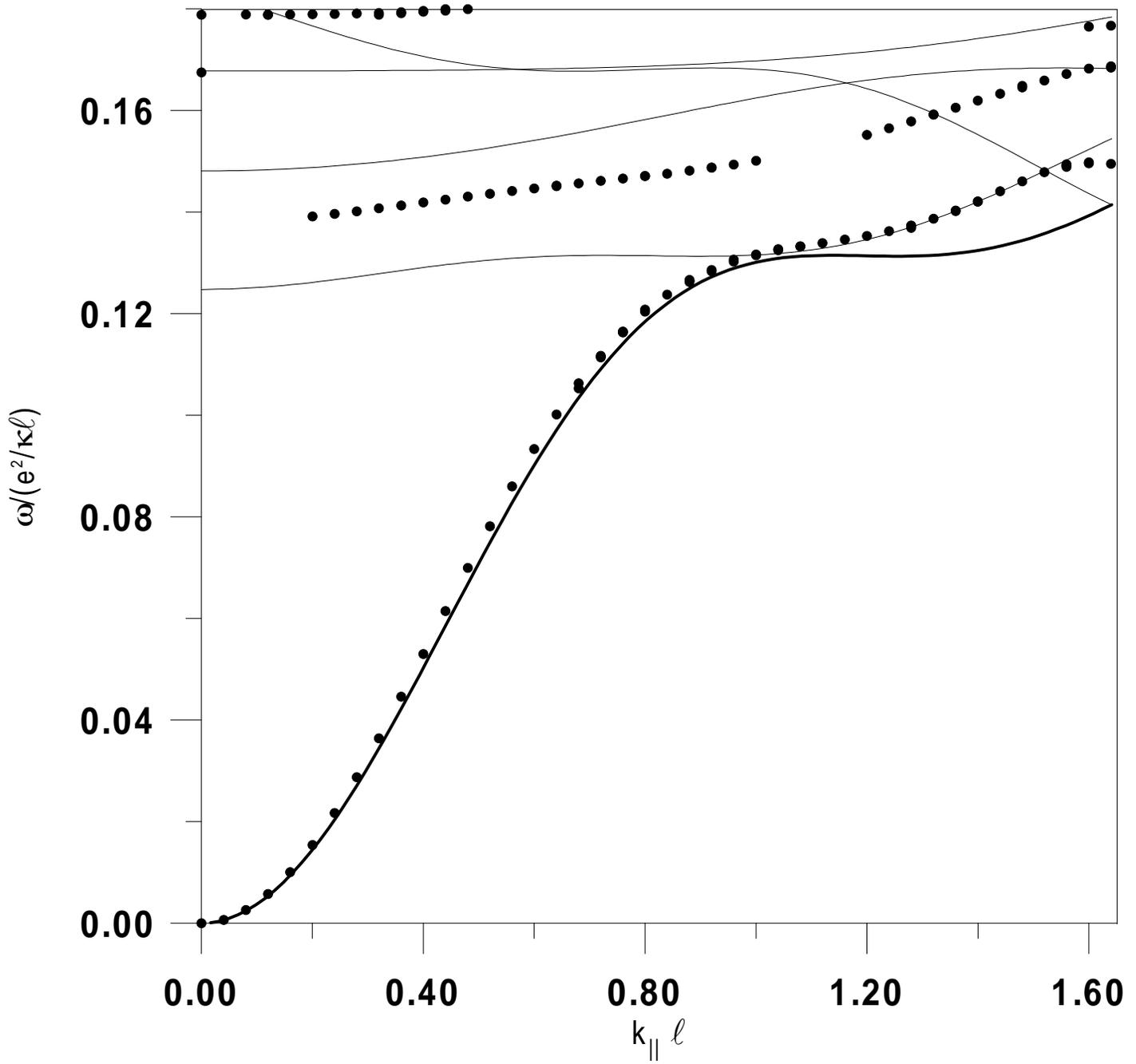
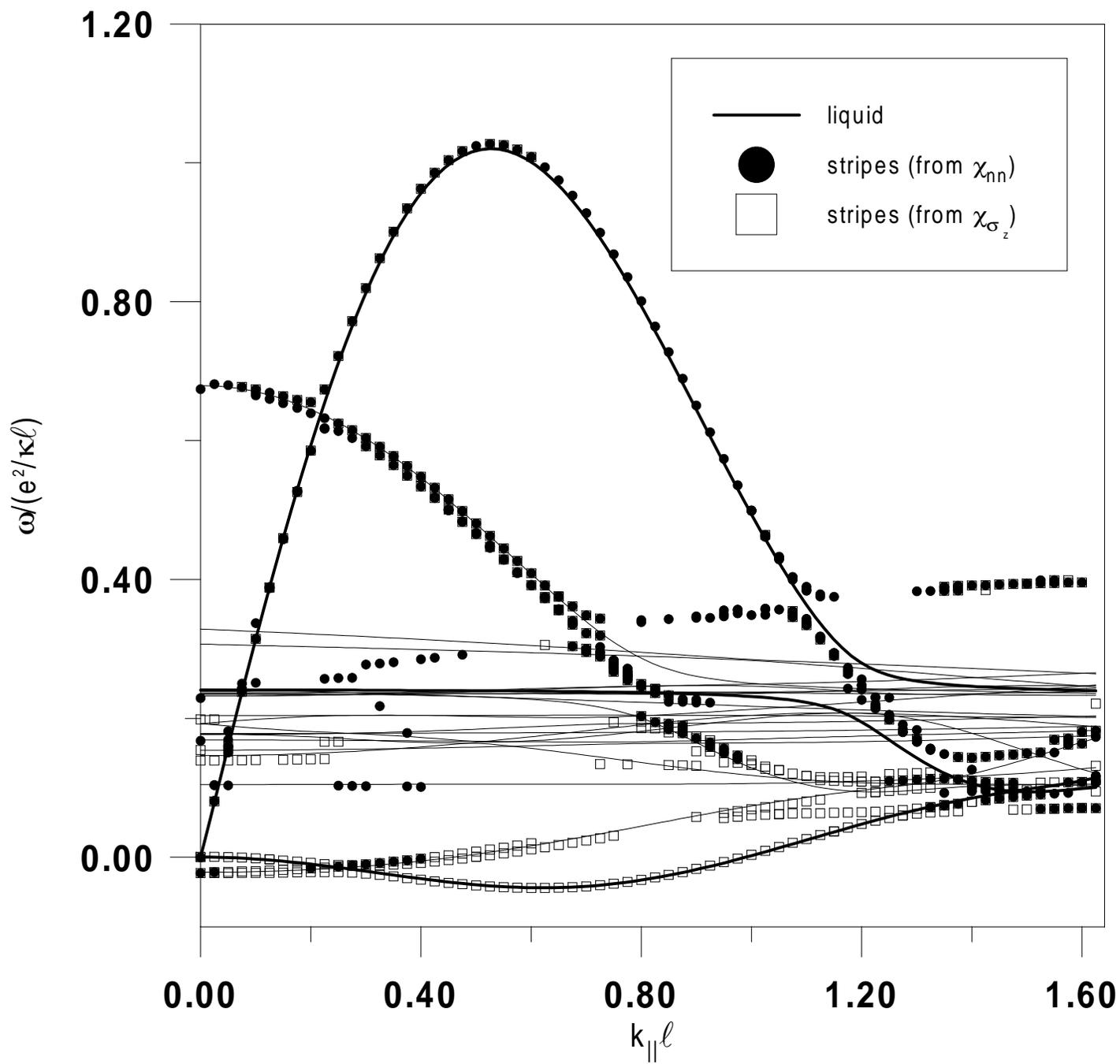
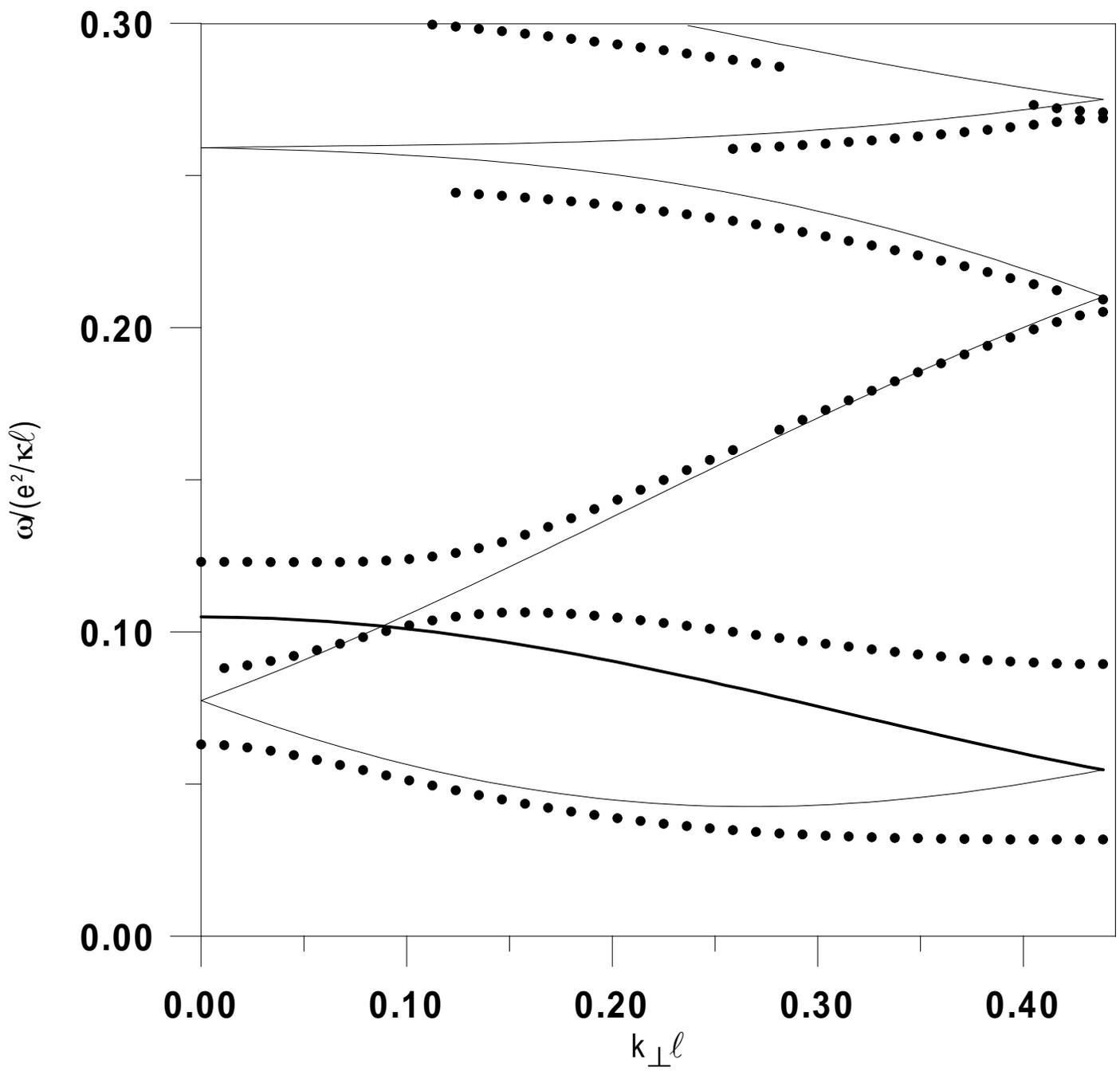
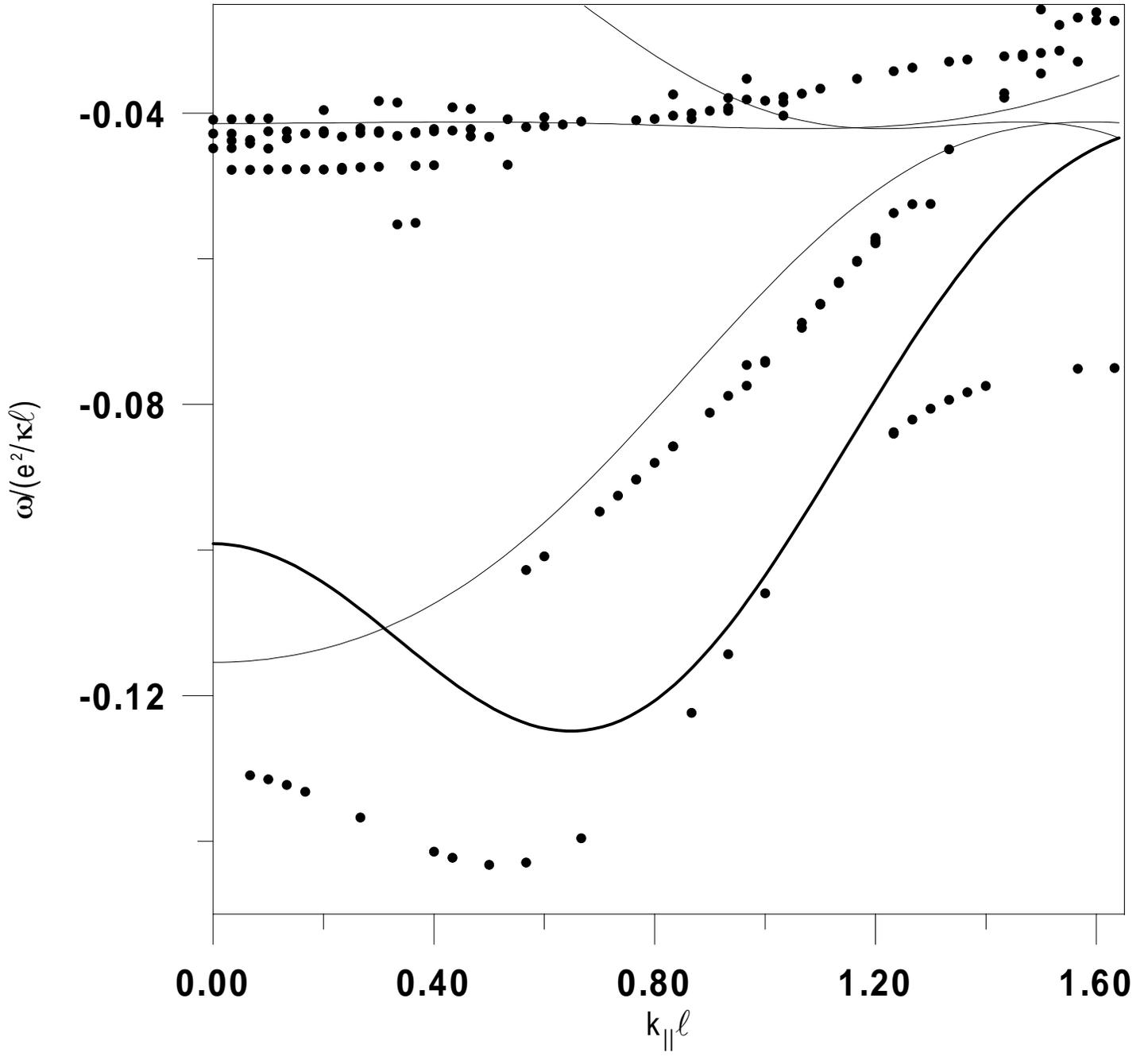
\end{document}